\documentclass[12pt]{article}

\usepackage{amssymb}
\usepackage{amsfonts}
\usepackage{amsmath,bbm}
\usepackage{mathtools}
\usepackage{appendix}
\usepackage{geometry}
\usepackage{scalefnt}
\usepackage[onehalfspacing]{setspace}
\usepackage{theorem}
\usepackage[pdftex]{graphicx}
\usepackage{subfigure}
\usepackage[pdfpagelabels]{hyperref}
\usepackage{filecontents}
\usepackage{ragged2e}
\usepackage{lipsum}
\usepackage{tikz}
\usetikzlibrary{positioning}
\usetikzlibrary{arrows}

\usepackage{color}
\usepackage{array}
\usepackage{multirow}
\usepackage{bigstrut}
\usepackage{units}

\usepackage{natbib}

\newtheorem{assumption}{Assumption}

\newtheorem{corollary}{Corollary}

\newtheorem{lemma}{Lemma}

\newtheorem{proposition}{Proposition}

\makeatletter
\renewcommand*{\@fnsymbol}[1]{\ensuremath{\ifcase#1\or *\or
 \mathsection\or \mathparagraph\or \|\or **\or \dagger\or \ddagger\or \dagger\dagger
 \or \ddagger\ddagger \else\@ctrerr\fi}}
\makeatother

\begin{document}

\pagenumbering{Alph}
\begin{titlepage}
\title{Free Riding in Networks\thanks{{\footnotesize We thank Nizar Allouch, Francis Bloch, Yann Bramoull\'{e}, Matt Elliott, Andrea Galeotti, Ben Golub, Timo Hiller, Jaromir Kovarik, Dotan Persitz, Javier Rivas, participants of the 2018 Network Science in Economics at Vanderbilt University, the Barcelona Summer Forum, the PSE Workshop on Dynamic Models of Interaction, BiNoMa and seminar participants at Haifa, Tel Aviv University, University of Kent and Saint Louis for their comments and suggestions. Some of the results in this paper previously circulated in working papers titled ``Collaboration in Endogenous Networks'' and ``The Private Provision of Public Goods in Endogenous Networks''. The usual disclaimers apply.}}}

\author{Markus Kinateder\thanks{{\footnotesize Departamento de Econom\'{i}a, Edificio Amigos, Universidad de Navarra, 31009
Pamplona, Spain; email: mkinateder@unav.es. Financial support from the Spanish Ministry of Education and Science through grant PGC2018-098131-B-I00 is gratefully acknowledged.}} \\
	Luca Paolo Merlino\thanks{Department of Economics, University of Antwerp, Prinsstraat 13, 2000 Antwerp, Belgium; email: LucaPaolo.Merlino@uantwerpen.be. Merlino gratefully acknowledges financial support from the CNRS and the Research Foundation Flanders (FWO) through grants G026619N and G029621N.}}
	
\maketitle
	\begin{abstract}
		\noindent Players allocate their budget to links, a local public good and a private good. A player links to free ride on others' public good provision. We derive sufficient conditions for the existence of a Nash equilibrium. In equilibrium, large contributors link to each other, while others link to them. Poorer players can be larger contributors if linking costs are sufficiently high. In large societies, free riding reduces inequality only in networks in which it is initially low; otherwise, richer players benefit more, as they can afford more links. Finally, we study the policy implications, deriving income redistribution that increases welfare and personalized prices that implement the efficient solution.
\end{abstract}
\thispagestyle{empty}
\end{titlepage}
\pagenumbering{arabic}

\section{Introduction}

Local public goods are important phenomena in our society. For example, individuals often acquire some information on different alternatives whose advantages they do not know either personally or through their peers. Since people benefit from their neighbors' investment, personal acquisition of information is a local public good. Information can be exchanged both via in-person communication \citep{udry,feick}, in online forums and on social platforms. For instance, on Twitter, people write posts and see the posts of their connections on their wall. Another example of local public goods is constituencies offering some goods or services that can be enjoyed also by the citizens of nearby constituencies, such as public parks, a pedestrian city center, cultural activities, and the like.

Importantly, the resulting pattern of spillovers can be represented as a directed network in which establishing a link to a player allows to free ride on her contribution.

In the aforementioned examples, not only contributing is costly, but also free riding, as individuals need to incur the opportunity cost of travelling to nearby constituencies or searching whom to follow and reading their posts on Twitter. These opportunity costs are different across individuals; for example, they are higher for those who receive higher wages, as their time is more valuable. Additionally, while the most important driver of contributors is a specific need for the public good, agents might have different resources and motivations to provide it. For example, some may post on Twitter because they have a lot of spare time, and others to become influencers. Some constituencies might invest a lot in culture because they are rich, others because of a pronounced interest of their inhabitants.

Heterogeneities in wages yield opposing distributional effects of free riding in networks: on the one hand, everyone can free ride, and this should reduce inequality. On the other hand, the richer might benefit more from free riding if this is costly. As a result, new technologies that changed the opportunity cost of free riding might have affected who benefits the most from free riding. This raises the question of how to design policies to increase welfare and redistribute resources. Indeed, the planner might want to subsidize the public good provision of certain individuals via monetary or non-monetary subsidies. For example, YouTube redistributes some of the revenue generated by user activity rewarding its contributors with many initiatives, ranging from awards to fan fests. 

To study these considerations, we propose a local public good game in which players allocate their time or budget to a public good, a private good, and to free ride on others' provision. Introducing a budget constraint captures that individuals have different amounts of resources and/or time to dedicate to public good provision. We also allow for heterogeneous preferences and relative prices of the private good with respect to the public good and linking. Thereby, we capture that individuals might have different motivations, derive a different utility and pay a different price for an identical bundle.

The main contributions of the paper are two. First, we derive sufficient conditions for equilibrium existence and characterize equilibrium networks in a general framework that embeds well-known models of private provision of (local) public goods. Second, we use these results to analyze the impact of the income distribution on inequality and derive policy implications.

Regarding the first contribution, we show that, if both the public and the private good are normal and neighbors' contributions sufficiently crowd out own-contribution, a Nash equilibrium in which players establish weakly profitable connections exists; furthermore, these networks are core-periphery graphs in which the largest contributors are linked among themselves. Indeed, any two contributors are connected if they produce more than the linking cost, as this frees resources for additional public or private good consumption. Moreover, everyone prefers to link to the largest contributors, as this generates the largest spillovers. Therefore, any strict Nash network is a nested split graph, in which periphery players have nested neighborhoods. This equilibrium characterization is consistent with empirical evidence.\footnote{For example, both on Twitter and in an online forum where users ask and respond to queries about Java, there is a core of players who actively collaborate with each other, while most users only free ride \citep{bastos,adamic}.} Our characterization is also robust to several extensions (Section \ref{extensions}).

The framework we propose allows us to revisit classical results in the literature on global public goods \citep{BBV,warr}. In our model, this corresponds to linking costs being zero. Then, when both the public and the private good are normal, a unique Nash equilibrium exists, in which total provision is neutral to income redistribution among contributors. If additionally players have identical preferences, the richest contribute and their contributions reduce inequality.

To the contrary, when the linking costs are larger, contributors are not necessarily the richest players even when preferences are homogeneous; moreover, poorer players can be better off than richer ones. To resolve the indeterminacy arising from multiple equilibria, we study inequality in large societies, for which we derive the so-called law of the few \citep{gg}, i.e., the proportion of players in the core converges to zero. In this case, we can analyze welfare focusing on free riders.

In particular, the impact of free riding in networks on inequality depends on the initial wealth distribution. When players have similar wealth levels, they can afford the same links and enjoy the same spillovers, thereby reducing inequality. If, to the contrary, a society is very unequal, some can afford more links and hence free ride more than others. As a result, in large societies, networks increase inequality when the economy is relatively unequal to begin with.

The mechanism we highlight implies that new technologies that make it cheaper to free ride on others will reduce inequality. 

Our model has important implications for policy interventions. In contrast to models with an exogenous network \citep{nizar}, here income redistribution affects linking incentives, giving the planner an additional constraint, but also an additional instrument, to improve welfare. Indeed, in an exogenous core-periphery graph, it is optimal to transfer income to core players. However, when players can change their links as a result of the intervention, it is better to transfer income to players who are not central because of limited resources, if they value the public good a lot, as this translates into a larger provision.

Beyond improving welfare, the efficient solution can be implemented if the social planner is able to charge personalized prices for the public good. This could be relevant for online social platforms, where public good provision of an individual can be subsidized via sponsorship, while other users can be taxed via digital services taxes that collect some of the revenue generated by user activity. For example, social platforms such as YouTube have a host of programs to reward largest contributors in order to incentivize the creation of new content, from awards to fan fests.

The novel policy implications of our model are due to following main innovations: players face a budget constraint, and both the network as well as the demand for the public good are endogenous. The latter assumption contrasts with existing models of games on endogenous networks that assume a fixed demand for the public good \citep{gg}. This paper is then the first to study free riding in endogenous networks under general preferences and heterogeneous budgets. 
When the budget is not relevant for the demand of the public good, players who value it more, provide more and are in more central positions in the network \citep{KM}. However, when the budget set matters, novel insights emerge on how poorer players can have more central positions or benefit more from free riding, thereby affecting both inequality and the policy implications.\footnote{\cite{leeuwenJEEA} introduce competition for status in such a framework. Status rents foster public good provision in the repeated game. Our model accommodates that spillovers increase consumption also in a static game by allowing for richer preferences. Our characterization reminds also of that of \cite{ramos}, but the two models are very different. Indeed, in their paper homogeneous players establish costly links to see other players' signals, and then play a beauty contest game. In \cite{dotan}, exogenously determined high types are in the core which provide low types in the periphery with cheap low-quality indirect connections.}

We allow for best replies to be non-linear in neighbors' provision and income, which is empirically relevant (e.g., \citealp{quadratic}). Previous analysis with similarly general preferences has been limited to games on fixed networks; e.g., \cite{nizar} shows that the equilibrium is unique if the public good is sufficiently normal. For endogenous networks, very strong normality not only does not yield uniqueness, but may even conflict with equilibrium existence. Here uniqueness arises for very low linking costs.\footnote{Most of the literature assumes only one good and linear best replies, both when the network is fixed 
\citep{keyplayer,brkran,brkrandam} or endogenous \citep{baetz,hiller}. In particular, \cite{ktz} study a dynamic potential game of strategic complements that predicts nested split graphs as stochastically stable in a myopic best reply dynamics. However, the nature of the strategic interaction is very different, as we focus on substitutes. We also allow for heterogeneous players and games that do not admit a potential. More importantly, we study a static and simultaneous game with rational players where nestedness obtains only in strict Nash equilibria.}

\cite{golub} characterize the Pareto efficient outcomes and Lindahl solution in a local public good game on an exogenously fixed network in terms of well-studied network statistics. In our paper, to the contrary, we are interested in understanding whether policies aimed at raising welfare would have the intended effects when players can change with whom to interact after the policy intervention.\footnote{\cite{belhaj2016} show that nested split graphs are also efficient networks in a network game with local complementarities. Our results on second-best implementation complements theirs, as we assume strategic substitutes.}

In our benchmark model, spillovers flow one-way, i.e., towards the player establishing the link.\footnote{Some papers study network formation with one-way flow of spillovers abstracting from strategic interactions on the network (e.g., \citealp{balagoyal}, \citealp{galeotti2006one}, \citealp{billand2008existence}).} To account for situations when the interaction is face-to-face, we extend our analysis to two-way flow of spillovers (see the Online Appendix). While most of our findings are robust, we also derive the sufficient condition for a stricter version of the law of the few.



The remainder of the paper is organized as follows. Section \ref{model} introduces the model. Section \ref{mainsec} provides results on the existence and characterization of equilibria. Section \ref{sec:homogeneous} focuses on economies with agents with homogeneous preferences to compare our results to those of \cite{BBV}. Section \ref{inequality} studies the impact of income (re)distribution and personalized prices on inequality. Section \ref{extensions} discusses some extensions. Section \ref{conclusions} concludes. All proofs are in the appendix.

\section{Model}\label{model}

We introduce a local public good game, in which players spend their budget on private good consumption, public good provision and connections.

\smallskip

\noindent \textbf{Players.} There is a set of players $N = \{1, ..., n\}$; $i$ denotes a typical player.

\smallskip
\noindent \textbf{Network.} We denote the directed network of social connections by $g$. Player $i$'s linking strategy is denoted by a row vector $g_i =( g_{i1}, ..., g_{in} ) \in G_i = \{0, 1\}^{n}$, where $g_{ii}=0$ and $g_{ij} \in \{0,1\}$ for all $i,j \in N$, $i \neq j$. We say that player $i$ links to player $j$ if $g_{ij}=1$. Then, $g=(g_1,...,g_n)^{T}$. Linking decisions are one-sided: the player proposing a link pays $k>0$ and the link is established. Let $N_i(g) = \{j \in N: g_{ij} = 1\}$ be the set of players to which $i$ links and $\eta_i(g) = \left\vert N_i(g) \right\vert$ the number of links that $i$ sponsors.

In a \textit{core-periphery graph}, there are two groups of players, the \textit{periphery} $\mathcal{P}(g)$ and the \textit{core} $\mathcal{C}(g)$, such that, \textit{(i)} for every $i,j \in \mathcal{P}(g)$, $g_{ij}=g_{ji}=0$, \textit{(ii)} for every $l,m \in \mathcal{C}(g)$, $g_{lm}=g_{ml}=1$, and \textit{(iii)} for any $i\in\mathcal{P}(g)$, there is $l\in\mathcal{C}(g)$ such that $g_{il}=1$. Hence, all links in the core are reciprocated. A \textit{complete core-periphery} network is such that $N_i(g)=\mathcal{C}(g)$ for all $i \in \mathcal{P}(g)$ and there are no isolated players. Nodes in $\mathcal{C}(g)$ are referred to as \textit{hubs}. We write $\mathcal{C}$ and $\mathcal{P}$ instead of $\mathcal{C}(g)$ and $\mathcal{P}(g)$, respectively, when no confusion arises. A core-periphery network with a single hub is referred to as a \textit{star}. A core-periphery network in which the sets of players' neighbors are nested is a \emph{nested split graph}. Formally, a nested split graph is a core-periphery graph where, if $\eta_j(g)\leq \eta_i(g)$, then $N_j(g) \subseteq N_i(g)$ for any $i,j\in \mathcal{P}(g)$.\footnote{We extend here in a natural way some graph theoretic notions usually defined on undirected networks to our model of directed network formation. The corresponding definitions for the two-way flow model are in the Online Appendix. In particular, the notion of core-periphery graph we use here is more specific than the ones defined on the closure of $g$ (see below), as it embeds that core players reciprocate links and periphery players link to the core.}

Denote by $\overline{g}$ the closure of $g$, such that $\overline{g}_{ij} = \max \{g_{ij}, g_{ji}\}$, for each $i,j \in N$; that is, each directed link in $g$ is replaced by an undirected one. Let $N_i(\overline{g}) = \{j \in N: \overline{g}_{ij} = 1\}$ be the set of players to which $i$ is linked in $\overline{g}$, and let $\eta_i(\overline{g}) = \left\vert N_i(\overline{g})\right\vert$ be $i$'s degree, i.e., the number of $i$'s neighbors in $\overline{g}$.

There is a path in $\overline{g}$ from $i$ to $j$ if either $\overline{g}_{ij}=1$, or there are $m$ different players $j_1,...,j_m$ distinct from $i$ and $j$, such that $\overline{g}_{ij_1}=\overline{g}_{j_1j_2}=...=\overline{g}_{j_m j}=1$. A \textit{component} of network $\overline{g}$ is a set of players such that there is a path connecting every two players in the set and no path to players outside the set. Define a cell $h$ of a nested split graph as the set of players $i \in N$ with $h = \eta_i(\overline{g})$ links.

 
\smallskip

\noindent \textbf{Consumption.} We denote by $x_i \in \mathcal{R}^+$ and $y_i \in \mathcal{R}^+$ the amount of public and private good acquired by player $i$, respectively, where $\mathcal{R}^+ \equiv [0, + \infty)$.

Given $g,$ we denote by $\overline{x}_{-i}= \sum_{j \in N} g_{ij} x_j$ player $i$'s spillovers and by $\overline{x}_i = x_i + \overline{x}_{-i}$ player $i$'s public good consumption, given by the sum of her provision and the spillovers she receives from her neighbors in network $g$. Hence, we assume that spillovers flow one-way, i.e., towards the player sponsoring the link. In Online Appendix A, we study the model with two-way flow of spillovers. 

\smallskip

\noindent \textbf{Strategies.} Player $i$'s set of strategies is $S_i = \mathcal{R}^+\times\mathcal{R}^+\times G_i$, and the set of all players' strategies is $S = \prod_{i \in N} S_i $. A strategy profile $s = (x,y, g) \in S$ specifies provision of the public good $x = (x_1,... ,x_n)$, consumption of the private good $y = (y_1,...,y_n)$, as well as links $g = (g_1, ... ,g_n)^T$ for each player. In order to emphasize player $i$'s role, $s$ is sometimes written as $(s_i, s_{-i})$, where $s_{-i}\in \prod_{j\neq i} S_j.$

\smallskip

\noindent \textbf{Payoffs.} Each player $i$ faces the following maximization problem:
\begin{eqnarray}\label{max}
\max_{(x_i,y_i,g_i)\in S_i} && U_i(\overline{x}_i,y_i) \\ \notag
\text{s.t.} & & x_i+p_i y_i+\eta_i(g) k=w_i,
\end{eqnarray}
where $x_i$ and $y_i$ are the respective amounts of public and private good personally acquired by $i$, while $\overline{x}_i$ is $i$'s public good consumption, $p_i > 0$ is the price of the private good paid by player $i$, $k > 0$ is the cost of linking and $w_i >0$ her wealth. We assume $U_i(\cdot,\cdot)$ is a twice continuously differentiable, strictly concave and increasing function in its arguments for each $i \in N$. Under these assumptions, there is a unique and non-negative optimal investment in the public and private good for every $i$, which depends on own wealth, the links sponsored and the spillovers received from neighbors. Denote player $i$'s optimal consumption in isolation by $(x_i^I, y_i^I)$.

By analyzing heterogeneous prices for the private good, we capture that the relative price of the private good with respect to that of the public good and the linking costs can differ across players. Additionally, some people might like the public good more than others. Hence, we allow players to have different preferences. 


The utility maximization problem can be rewritten with player $i$ choosing her local public good consumption $\overline{x}_i$, rather than her public good provision $x_i$:
\begin{eqnarray}\label{max2}
\max_{( \overline{x}_i,y_i,g_i)\in S_i} && U_i(\overline{x}_i,y_i) \\ \notag
\text{s.t.} && \overline{x}_i+p_i y_i= w_i-\eta_i(g) k+ \overline{x}_{-i}, \text{ and } \overline{x}_i \geq \overline{x}_{-i}.
\end{eqnarray}
Ignoring the inequality constraint, we can express player $i$'s demand function for the public good as $\overline{x}_i=\gamma_i(w_i-\eta_i(g) k+ \overline{x}_{-i})$, where $\gamma_i:\mathcal{R}\rightarrow\mathcal{R}$ is a continuous function. We call $\overline{w}_i(g)=w_i-\eta_i(g) k+ \overline{x}_{-i}$ player $i$'s \textit{net social income} and $\gamma_i$ $i$'s Engel curve. Hence, $x_i=\overline{x}_i-\overline{x}_{-i}=\max\{\gamma_i(\overline{w}_i(g))-\overline{x}_{-i},0\}$. Since the network is endogenous, player $i$'s demand for the public good is net of the budget she spends on links.

We assume that both the public and the private good are normal for all players.\footnote{Note that, when $i$'s links change, $i$'s net social income typically changes discontinuously. However, $\gamma$ \textit{per-se} is a continuously differentiable function and jumps in net social income do not pose a problem for our analysis.}
\begin{assumption}\label{normality}
For each $i\in N$, $\gamma_i(w)$ is continuously differentiable with respect to $w$ with derivative $\gamma_i^{\prime}\in[0,1]$.
\end{assumption}
Let us stress that this assumption is mild and commonly made in public good games. In particular, our model embeds the global public good game of \cite{BBV} as $k\rightarrow 0$. Then, contributors form a core, periphery players link to all contributors and there are no spillovers among inactive players.\footnote{When $k=0$, a complete network with links among inactive players is also an equilibrium. Note that this is also a core-periphery graph, and still there are no spillovers among inactive players. Also note that we allow for $\gamma_i^{\prime}\in\{0,1\}$. As in \cite{BBV}, these extremes do not pose a problem for existence, but they do for uniqueness of the equilibrium (Corollary \ref{ex-bbv}).}

\smallskip

\noindent \textbf{Equilibrium.} A strategy profile $s^\ast = (x^\ast, y^\ast, g^\ast)$ is a \emph{Nash equilibrium} if for all $i \in N$, $s_i^{\ast}$ is a solution to the maximization problem \eqref{max} given $s_{-i}^\ast$.

To characterize the equilibrium networks, we will use two refinements. We say a Nash equilibrium is \emph{sociable} if any player $i$ who is indifferent between establishing a link or not, establishes this link. More formally, a Nash equilibrium $(x^\ast,g^\ast)$ is sociable if, when there is $(x_i^\prime,y_i^\prime)$ for $i\in N$ such that $U_i(\bar{x}^\ast,y^\ast)=U_i(\bar{x}_i^\prime,y^\prime_i|\bar{x}^\ast_{-i},y^\ast_{-i})$, then, for any $j\in N\setminus \{i\}$, $g^{\ast}_{ij}\geq g^{\prime}_{ij}$, with strict inequality for some $j$. 
This refinement allows us to discard equilibrium networks that are not robust to small changes in players' wealth and preferences.

An equilibrium is \emph{strict} if no player can unilaterally change her strategy without reducing her payoff. Hence, any strict equilibrium is sociable.


\smallskip

\noindent \textbf{Social Welfare.} We define social welfare as the sum of individual payoffs.

\smallskip

\noindent \textbf{Discussion and interpretation.} The presence of a budget set and that players pay to free ride are the key innovations of our model. The budget constraint stems from players having limited wealth or time to devote to these activities; links are one-sided and the flow of spillovers is one-way. These assumptions capture situations in which one needs to exert some effort to enjoy the public good provided by others, such as reading a post on Twitter. In this example, players decide how to allocate their time between working and leisure. During leisure, they either collect information directly and post it on Twitter, or read the posts of the users they follow. Heterogeneous wages result in heterogeneous relative prices for the private good.

This generates novel trade-offs between using resources for free riding by linking, public good provision and private good consumption. Most importantly, a higher net social income translates into a higher demand for the public good whenever $\gamma_i^{\prime}>0$.

In the next sections, we characterize the sociable Nash equilibria of this game, and discuss who the largest contributors are. Then, we will discuss how degree of heterogeneity in the initial endowment or in wages (if the constraint is a time constraint) shapes the distributional effect of free riding in networks. We will then analyze the policy implications for income redistribution and the design of personalized prices.

\section{Main Analysis}\label{mainsec}

Let us first discuss two lemmas that highlight the novel aspects of our model. First, differently from \cite{gg} and \cite{KM}, spillovers do not completely crowd out own contributions if $\gamma_i^{\prime}>0$. As we show later, this has an impact on the existence and characterization of equilibrium, and in particular, on who the largest contributors are.
\begin{lemma}\label{contribution}
If Assumption \ref{normality} holds, contributions are decreasing in spillovers.
\end{lemma}
This result follows immediately from the fact that $\overline{x}_{i}=\max\{\gamma_i( \overline{w}(g)),0\}$ and $\gamma_i^{\prime}\in[0,1]$ (Assumption \ref{normality}), so that demand for the public good is endogenous and increasing in net social income.

Second, despite heterogeneity in preferences, the introduction of a budget constraint implies that largest contributors are linked.
\begin{lemma}\label{almost_core}
Given a Nash equilibrium, suppose there are players $i,j\in N$ with $x^{\ast}_i, x^{\ast}_j>k$. Then, if Assumption \ref{normality} holds, $g_{ij}^{\ast} = 1$.
\end{lemma}
When the budget set is irrelevant \citep{KM}, the gains from a connection determines who links with whom; here instead, all players who produce more than the linking cost are linked to each other even if they can have very different gains from being connected. Indeed, if two such players were not neighbors, they could increase their net social income by free riding on each other's provision. This increases their demand for the public good if it is normal.

This novel insight drives the following proposition.
\begin{proposition}\label{characterization}
If $k>\overline{k}=max_{i\in N}\{x_i^I\}$, the unique equilibrium network is empty. If $k\leq \overline{k}$ and Assumption \ref{normality} holds, any sociable Nash equilibrium network is a core-periphery graph, while any strict Nash equilibrium network is a nested split graph.
\end{proposition}
This sharp characterization emerges from two equilibrium requirements. First, by Lemma \ref{almost_core}, players who produce strictly more than the linking cost $k$ form a core of interconnected players. The refinement of sociable Nash equilibrium ensures that the largest contributors form a core. Indeed, if there were several players providing an amount identical to $k$, then free riders would be indifferent between linking and providing an amount $k$ of the public good themselves. Networks without a core could be sustained in equilibrium due to these kinds of indifference. Second, conditional on linking, a player links to the largest contributors. Together with the existence of the core, this generates core-periphery structures.

When players are indifferent between linking to players providing the same amount of public good, but do not want to link to all of them, they could link to different ones. The refinement of strict Nash equilibrium ensures that nested split graphs emerge in such situations, as everybody links to larger contributors first.\footnote{Note that core players in a strict Nash equilibrium can have identical provisions, as long as no periphery player links to only one of them.}

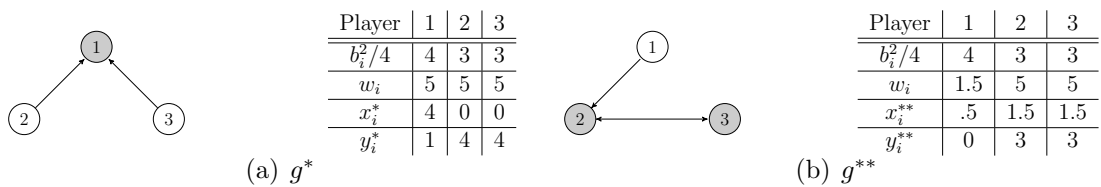
\begin{figure}[htbp!]
\begin{center}
\begin{minipage}{.48\linewidth}
\subfigure[$g^\ast$]{\label{1aej}
\begin{minipage}{.5\linewidth}
\resizebox{2.5cm}{!} {
	\begin{tikzpicture}[-,>=stealth',shorten >=1pt,auto,node distance=2.4cm and 2.8cm,
         main node/.style={circle,draw}]

 \node[main node,fill=black!20] (1) {1};
 \node[main node] (2) [below left of=1] {2};
 \node[main node] (3) [below right of=1] {3};

 \path[every node/.style={font=\sffamily\small}]
 (2) edge [->] node [right] {} (1)
 (3) edge [->] node [right] {} (1);
\end{tikzpicture}
	}
	\end{minipage}
	\begin{minipage}{.5\linewidth}
	\begin{center}
\resizebox{2.7cm}{!} {
\begin{tabular}{c|c|c|c}
 Player & 1 & 2 & 3 \\
\hline \hline $b_i^2/4$& 4 & 3 & 3 \\
\hline $w_i$& 5 & 5 & 5 \\ 
\hline $x_i^{\ast}$& 4 & 0 & 0 \\
\hline $y_i^{\ast}$& 1 & 4 & 4
\end{tabular}
}
\end{center}
\end{minipage}
	}
	
\end{minipage}
\begin{minipage}{.48\linewidth}
\subfigure[$g^{\ast\ast}$]{\label{2aej}
\begin{minipage}{.5\linewidth}
\resizebox{2.5cm}{!} {
	\begin{tikzpicture}[-,>=stealth',shorten >=1pt,auto,node distance=2.4cm and 2.8cm,
     main node/.style={circle,draw}]

 \node[main node] (1) {1};
 \node[main node,fill=black!20] (2) [below left of=1] {2};
 \node[main node,fill=black!20] (3) [below right of=1] {3};

 \path[every node/.style={font=\sffamily\small}]
 (2) edge [<-] node [right] {} (1)
 (2) edge [<->] node [right] {} (3)
 ;
\end{tikzpicture}
	}
	\end{minipage}
	\begin{minipage}{.5\linewidth}
	\begin{center}
\resizebox{3.4cm}{!} {
\begin{tabular}{c|c|c|c}
 Player & 1 & 2 & 3 \\
\hline \hline $b_i^2/4$& 4 & 3 & 3 \\
\hline $w_i$& 1.5 & 5 & 5 \\
\hline $x_i^{\ast\ast}$ & .5 & 1.5 & 1.5 \\
\hline $y_i^{\ast\ast}$ & 0 & 3 & 3
\end{tabular}
}
\end{center}
\end{minipage}
	}
\end{minipage}
\caption{Two examples with $U_i(\overline{x}_i,y_i) =\sqrt{b_i\sqrt{\overline{x}_i}+y_i}$, $b_1=4$, $b_2=b_3=2\sqrt{3}$, $k=1$, and $p_i=1$ for all $i\in N$. Players in the core are shaded in gray.}\label{comparison_aej}
\end{center}
\end{figure}

The limited role of preferences in Proposition \ref{characterization} is due to the main novelty of our model: the introduction of the budget constraint. Indeed, when it does not constrain the demand for the public good, players who value the public good more have more links, and produce more if they are in the core (Theorem 2, \citealp{KM}). This happens because players with a high valuation not only have a larger demand, but also have larger gains from a connection, as in Figure \ref{1aej}. In the example, player $i$'s demand of the public good is $\min\{b_i^2/4,w_i\}$. Since in \ref{1aej} $w_i>b_i^2/4$ for all players, those who value the public good more provide more. In panel \ref{2aej} instead $w_1 < b_1^2/4$, player $1$, who values the public good the most, is in the periphery because she is too poor to contribute enough to be in the core of a sociable equilibrium network. A similar phenomenon can arise because of heterogeneity in the price of the private good.\footnote{For example, the network in Figure \ref{1aej} is an equilibrium also when $U_i(\overline{x}_i,y_i) =(\sqrt{\overline{x}}+\sqrt{y_i})^2$ for all $i\in N$, $k=2$, $w=(10,6,6)$, $p_1=1$ and $p_2=p_3=1.5$. In that case, $x^\ast=(3,0,0)$, even if $2$ and $3$ face a higher price for the private good and this is a substitute of the public good.} As we show later, taking into account these phenomena is particularly relevant in the design of policy interventions.


Since we do not have linear best replies, the few conditions we impose to derive Proposition \ref{characterization} do not ensure equilibrium existence for two reasons: on the one hand, free riding decreases contribution (Lemma \ref{contribution}); on the other, it increases the demand for the public good. To see why this matters, suppose there is a player $i$ in the periphery who produces enough public good for other players to profitably link to $i$.\footnote{Note that, if $i$ is best-responding, $i$ is already linked to every player in the core; otherwise, doing so would constitute a profitable deviation.} Clearly this is no equilibrium, so consider moving $i$ in the core. In that case, players who remain in the periphery free ride also on $i$, thereby increasing their demand for the public good. Hence, if their own provision is not reduced enough, they might now themselves produce more than $k$, 
which is also no equilibrium. In sum, if players' Engel curves are very steep, 
it might be possible to construct cycles of deviations where players might produce too much when in the periphery, but not enough when in the core.

Yet, it is sufficient for equilibrium existence to bound the slope of the Engel curve, $\gamma^{\prime}$. This bounds the increase in players' demand for the public good when they free ride on others. Moreover, it ensures that spillovers substantially crowd out contributions of free riders.
\begin{proposition}\label{existence}
If Assumption \ref{normality} holds, there is $\overline{\gamma}\leq 1$ such that a sociable Nash equilibrium exists if $\gamma_i^{\prime} \leq \overline{\gamma}$ for all $i\in N$ for whom $x_i^I \geq k$.
\end{proposition}
Proposition \ref{existence} complements existing results for exogenous networks, which show existence of a unique equilibrium when $\gamma_i^{\prime}$ is sufficiently large \citep{nizar}. The game we study instead has potentially a very large set of equilibria.

\begin{figure}[htbp!]
\begin{center}
\begin{minipage}{.34\linewidth}
\centering
\resizebox{2.5cm}{!}{
\begin{tikzpicture}[-,>=stealth',shorten >=1pt,auto,node distance=2.4cm and 2.8cm,
     main node/.style={circle,draw}]

 \node[main node,fill=black!20] (1) {1};
 \node[main node] (2) [below right of=1] {2}; 
 \node[main node] (3) [below left of=1] {3};

 \path[every node/.style={font=\sffamily\small}]
 (2) edge [->] node [right] {} (1)
 (3) edge [->] node [right] {} (1);
\end{tikzpicture}
	}
\end{minipage}
\begin{minipage}{.32\linewidth}
\centering
\resizebox{2.5cm}{!}{
	\begin{tikzpicture}[-,>=stealth',shorten >=1pt,auto,node distance=2.4cm and 2.8cm,
     main node/.style={circle,draw}]

 \node[main node,fill=black!20] (1) {1};
 \node[main node,fill=black!20] (2) [below right of=1] {2}; 
 \node[main node] (3) [below left of=1] {3};

 \path[every node/.style={font=\sffamily\small}]
 (3) edge [->] node [right] {} (1)
 (2) edge [<->] node [right] {} (1)
 (3) edge [->] node [right] {} (2);
\end{tikzpicture}
	}
\end{minipage}
\begin{minipage}{.32\linewidth}
\centering
\resizebox{2.5cm}{!} {
	\begin{tikzpicture}[-,>=stealth',shorten >=1pt,auto,node distance=2.4cm and 2.8cm,
     main node/.style={circle,draw}]

 \node[main node] (1) {1};
 \node[main node,fill=black!20] (2) [below right of=1] {2}; 
 \node[main node,fill=black!20] (3) [below left of=1] {3};

 \path[every node/.style={font=\sffamily\small}]
 (2) edge [<-] node [right] {} (1)
 (2) edge [<->] node [right] {} (3)
 (3) edge [<-] node [right] {} (1);
\end{tikzpicture}
	}
\end{minipage}

\vspace{.3cm}

\begin{minipage}{.32\linewidth}
\centering
\subfigure[$g^\ast$]{\label{1eq}
\resizebox{3cm}{!}{
\begin{tabular}{c|c|c|c}
 Player & 1& 2 & 3 \\
\hline $w_i$& 10& 8 & 8 \\ 
\hline $x_i^{\ast}$& 9.9 & 3.91 & 3.91 \\
\hline $y_i^{\ast}$& .1 & .14 & .14
\end{tabular}
}}
\end{minipage}
\begin{minipage}{.32\linewidth}
\centering
\subfigure[$g^{\ast\ast}$]{\label{2eq}
\resizebox{3cm}{!} {
\begin{tabular}{c|c|c|c}
 Player & 1& 2 & 3 \\
\hline $w_i$& 10& 8 & 8 \\
\hline $x_i^{\ast\ast}$& 5.95 & 3.95 & 0 \\
\hline $y_i^{\ast\ast}$& .1 & .1 & .1
\end{tabular}
}}
\end{minipage}
\begin{minipage}{.32\linewidth}
\centering
\subfigure[$g^{\ast\ast\ast}$]{\label{3eq}
\resizebox{3cm}{!} {
\begin{tabular}{c|c|c|c}
 Player & 1& 2 & 3 \\
\hline $w_i$& 10& 8 & 8 \\ 
\hline $x_i^{\ast\ast\ast}$& 2 & 3.97 & 3.97 \\
\hline $y_i^{\ast\ast\ast}$& .1 & .08 & .08
\end{tabular}
}}
\end{minipage}

\caption{Multiple equilibria for $k=3.95$, $U_i(\overline{x}_i,y_i) = \overline{x}_i^a y_i^{1-a}$, $\gamma_i^{\prime}(\overline{w}_i)=a=.99$ and $p_i=1$ for all $i\in N$. Players in the core are shaded in gray.}\label{multiple_equilibria}
\end{center}
\end{figure}
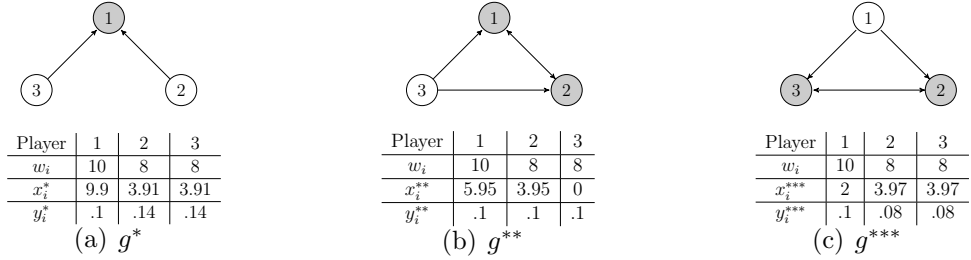

However, a very large $\gamma^\prime$ does not guarantee uniqueness.\footnote{As usual, we define uniqueness up to permutations in the labels of otherwise identical players.} Indeed, Figure \ref{multiple_equilibria} presents an economy with three players with linear Engel curves with slope $.99$; while the assumption for uniqueness of \cite{nizar} is satisfied, all three networks are an equilibrium. Rather, in our framework the equilibrium is unique when the linking cost is sufficiently small, as then the ranking of players' demand for the public good is not affected by their linking strategies. In other words, if $k$ is sufficiently small, free riding opportunities are the same for all players, so that the set of large contributors is unique as in \cite{BBV}.
\footnote{
In the example of Figure \ref{multiple_equilibria}, $g^{\ast\ast}$ is the unique equilibrium if $2.63<k<3.93$, and the complete network if $k\leq 2.63$.}

Furthermore, the unique equilibrium on a fixed network is not necessarily an equilibrium when the network is endogenous, because it might predict periphery players to be large contributors; this is impossible when the network is endogenous.\footnote{Consider the third example in Figure 3 in \citet[p.~541]{nizar}. The equilibrium is a star where the hub free rides on the contributions of the periphery. This is no equilibrium with endogenous links as contributors would free ride on each others' effort, thereby reducing their provision.}

This result is stated in the following corollary.\footnote{We focus on sociable equilibria in the following corollary because there are many equilibria when $k=0$ as individuals are indifferent to delete or add links to inactive players. Note also that there is trivially a unique equilibrium when the linking cost is very high: a star with the largest contributor as hub, and eventually, the empty network as the linking cost increases further.}
\begin{corollary}\label{ex-bbv}
There exists $\tilde{k}$ such that if $0\leq k<\tilde{k}$ and $\gamma_i^{\prime} \in (0,1)$ for all $i\in N$, there always exists a unique sociable Nash equilibrium entailing a core-periphery network.
\end{corollary}
Intuitively, the game we study has potentially a very large set of equilibria because different positions in the network imply very different net social incomes, thereby affecting demand and provision of the public good. In particular, a player's budget available for provision decreases as more links are established. However, as the linking costs decrease, a player's demand is less sensitive to the number of links established. Hence, in the limit as $k\rightarrow 0$, the classical uniqueness result of \cite{BBV} is reestablished.

\subsection{Large Societies}\label{large_soc}

The law of the few predicts that, as the number of players increases, the proportion of active players in non-empty strict equilibrium networks goes to zero \citep{gg}. In our model, this holds only for the proportion of players in the core. Let $\omega$ be the maximal wealth that players possess.
\begin{proposition}\label{lotf}
Suppose Assumption \ref{normality} holds. Given a scalar $\omega > 0$, in every sociable Nash equilibrium, $\lim_{n\rightarrow\infty}|\mathcal{C}(\overline{g}^{\ast})|/n=0$, if $w_i \leq \omega$ for all $i \in N$ as $n \rightarrow \infty$.
\end{proposition}
Intuitively, if there were many active players providing more than $k$ of the public good, they would all be connected. However, this is not possible as we impose the restriction that all players have a finite budget bounded by $\omega$. In other words, most of them need to free ride.

Proposition \ref{lotf} allows us to focus on periphery players to understand the welfare effects of networks. Indeed, the welfare effects on core players are less clear, since different players are in the core in different equilibria.

\section{Homogeneous Preferences and Prices}\label{sec:homogeneous}

We next study an economy in which players have identical preferences and face an identical price for the private good.
\begin{assumption}\label{homogeneous}
Every player has the same preferences and pays the same price for the private good.
\end{assumption}
When preferences are homogeneous and the linking costs are sufficiently low, as in \cite{BBV}, there is a unique threshold of wealth such that players whose wealth is above this threshold are in the core and consume the same amount of the private as well as of the public good, while the others are in the periphery and free ride on the contributors. When linking costs are higher, but the budget constraint is not binding for public good provision, differences in wealth instead are irrelevant. In that case, players with the highest demand of the public good are in the core, while the others are in the periphery or isolated \citep{KM}.

As the next proposition shows, results are more nuanced in our model, as poorer players can be in the core if richer players free ride a lot, thereby reducing their provision until it is not profitable to link to them.
\begin{proposition}\label{threshold}
If Assumptions \ref{normality} and \ref{homogeneous} hold, a sociable Nash equilibrium $(x^\ast,y^\ast,g^\ast)$ always exists. Furthermore:\\
\textit{(i)} For any $i,j \in \mathcal{C}(g^\ast),$ $\overline{x}_i^* = \overline{x}_j^*$ and $y_i^* = y_j^*$, while $x_i^* > x_j^*$ if and only if $w_i > w_j.$\\
\textit{(ii)} If there is $j \in \mathcal{P}(g^*)$ such that $g_{ji}^{\ast}=1$ for all $i \in \mathcal{C}(g^*)$, then $\overline{x}_j^* \geq \overline{x}_i^*$ for all $i \in \mathcal{C}(g^*)$.\\
\textit{(iii)} For any $i,j\in \mathcal{P}(g^*)$ such that $N_i(g^*) = N_j(g^*)$, $\overline{x}_i^\ast \geq \overline{x}_j^\ast$, $x_i^\ast \geq x_j^\ast$ and $y_i^\ast \geq y_j^\ast$ if and only if $w_i\geq w_j$.\\
\textit{(iv)} Take $i, j \in \mathcal{P}(g^*)$. If $w_i\geq w_j$, then $N_i(g^*) \geq N_j(g^*)$.\\
\textit{(v)} There is $w^\prime$ such that $i\in \mathcal{P}(g^*)$ if $w_i\leq w^\prime$.
\end{proposition}
Part \textit{(i)} shows that in our model only players in the core consume the same bundle, while this is not true for players who contribute less. Part \textit{(ii)} shows that a periphery player who connects to all players in the core consumes more than them, while she contributes less. Indeed, such a player free rides the most, while her wealth can be larger or smaller than that of core players. Larger spillovers increase her net social income which, due to normality, translates into a higher demand for the public good. This does not arise in models without income effect \citep{KM}.\footnote{This result is not due to the assumption of one-way flow of spillovers; indeed, we show in the Online Appendix that it also holds under the alternative assumption of two-way flow. A similar effect arises when the largest contributors are not connected on a fixed network \citep{brkran,nizar}. However, these configurations are no equilibrium when the network is endogenous, as these contributors would like to free ride on each other.}

While the monotonicity of provision in wealth holds for all players in global public good games, parts \textit{(iii)} and \textit{(iv)} show that here this holds among players with the same neighbors. Hence, it depends on who free rides on whom. Similarly, within the periphery, a player with more neighbors has to be richer to afford more links, free ride more and thereby possibly contribute less. Indeed, periphery players who sponsor less links may provide more public good: as they have less resources to sponsor links, they free ride less, which increases their contributions.

A consequence of part \textit{(iv)} is that periphery players tend to afford a different number of links the larger is wealth inequality. Therefore, the number of cells, that is, the sets of players having the same links in $\overline{g}$, weakly increases when wealth dispersion increases. In particular, complete core-periphery graphs emerge when players have similar wealth levels, since then all players can afford the same links.
\begin{corollary}
If Assumptions 1 and 2 hold, let $\tilde{w} = \min_{j \in \mathcal{P}} w_j$. Then, in any sociable Nash equilibrium, as $\tilde{w}$ decreases, the number of cells increases.
\end{corollary}

Last, but not least, part \textit{(v)} shows that we can identify those players who are poor enough so that they are always in the periphery. However, there is neither a threshold of wealth above which players are contributors nor one above which they belong to the core. Hence, while the richest players are the largest contributors when the linking costs are sufficiently low (Corollary \ref{ex-bbv}), when $k$ increases, poorer players can be in more central positions.



\begin{figure}[htbp!]
\begin{center}
\begin{minipage}{.48\linewidth}
\subfigure[$g^\ast$]{\label{1eqw}
\resizebox{2.8cm}{!}{
	\begin{tikzpicture}[-,>=stealth',shorten >=1pt,auto,node distance=2cm and 2cm,
          main node/.style={circle,draw}]

 \node[main node,fill=black!20] (1) {1};
 \node[main node,fill=black!20] (2) [right of=1, xshift=20mm] {2}; 
 \node[main node] (3) [below of=1] {3};
 \node[main node] (4) [below of=2] {4};

 \path[every node/.style={font=\sffamily\small}]
  (2) edge [<->] node [right] {} (1)
  (3) edge [->] node [right] {} (1)
  (3) edge [->] node [right] {} (2)
  (4) edge [->] node [right] {} (1);
\end{tikzpicture}
	}
	
\resizebox{4cm}{!} {
\begin{tabular}{c|c|c|c|c}
 Player & 1& 2 & 3 &4\\ 
 \hline $w_i$& 10 & 9 & 8 & 4 \\
\hline \rule{0pt}{12pt} $x_i^{\ast}$& $3.\overline{3}$ & $2.\overline{3}$ & $.1\overline{6}$ & $0$ \\
\hline \rule{0pt}{12pt}
$y_i^{\ast}$& $5.\overline{6}$ & $5.\overline{6}$ & $5.8\overline{3}$ & $3$
\end{tabular}
}
	}
\end{minipage}
\begin{minipage}{.48\linewidth}
\subfigure[$g^{\ast\ast}$]{\label{2eqw}
\resizebox{2.8cm}{!}{
	\begin{tikzpicture}[-,>=stealth',shorten >=1pt,auto,node distance=2cm and 2cm,
          main node/.style={circle,draw}]

 \node[main node,fill=black!20] (1) {1};
 \node[main node] (2) [right of=1, xshift=20mm] {2}; 
 \node[main node,fill=black!20] (3) [below of=1] {3};
 \node[main node] (4) [below of=2] {4};

 \path[every node/.style={font=\sffamily\small}]
  (2) edge [->] node [right] {} (3)
  (1) edge [<->] node [right] {} (3)
  (1) edge [<-] node [right] {} (2)
  (4) edge [->] node [right] {} (1);
\end{tikzpicture}
	}
\resizebox{4cm}{!} {
\begin{tabular}{c|c|c|c|c}
 Player & 1& 2 & 3 &4\\ 
 \hline $w_i$& 10 & 9 & 8 & 4 \\
\hline \rule{0pt}{12pt} $x_i^{\ast\ast}$& $3.\overline{6}$ & $.8\overline{3}$ & $1.\overline{6}$ & $0$ \\
\hline \rule{0pt}{12pt} $y_i^{\ast\ast}$& $5.\overline{3}$ & $6.1\overline{6}$ & $5.\overline{3}$ & $3$
\end{tabular}
}
	}
\end{minipage}

\caption{Comparing welfare between two equilibria for $k=1$, $U_i(\overline{x}_i,y_i) =\sqrt{\overline{x}_i y_i}$ and $p_i=1$ for all $i\in N$. Players in the core are shaded in gray.}\label{multiple_welfare}
\end{center}
\end{figure}
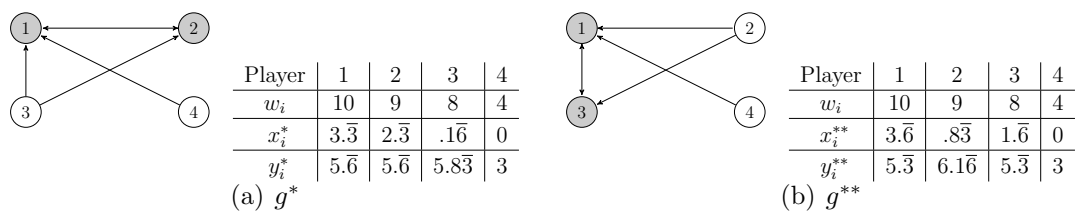

When multiple equilibria exist, it is interesting to understand how they compare in terms of welfare. To fix ideas, consider Figure \ref{multiple_welfare}. In this example, there are two equilibria: in \ref{1eqw}, the richest players are in the core; in \ref{2eqw}, player $3$ is in the core, despite of being poorer than $2$. Total public good provision is higher in \ref{1eqw}. However, in \ref{2eqw}, player $1$ contributes more to the public good to compensate the lower contribution of $3$. As a result, player $4$, who can afford only one link, is better off in \ref{2eqw}. Additionally, if there are at least three players like $4$ (with a wealth of $4$ and linking only to $1$), equilibria like \ref{2eqw} are associated with a higher welfare.

\noindent This example highlights a general result: if players have similar wealth---so that they can afford the same links,---and the linking cost is sufficiently high---so that adding players to the core is too costly,---an equilibrium with few and richer players in the core yields a higher welfare. Denote this equilibrium by $(x^h,y^h,g^h)$.\footnote{This equilibrium is the one we constructed in the proof of Proposition \ref{threshold} to show existence.} Then:
\begin{corollary}\label{richer}
Suppose that there exists a sociable equilibrium $(x^\prime,y^\prime,g^\prime)$ different from $(x^h,y^h,g^h)$. If $\gamma^{\prime}\in(0,1)$ and Assumption 2 holds, there exist $\omega>0$ and $K$ such that welfare is higher in $g^h$ than in $g^\prime$ if $\max_{i\in N}{w_i}-\min_{i\in N}{w_i}<\omega$ and $k>K$.
\end{corollary}
To conclude, we have just shown how, with homogeneous preferences, our model yields precise predictions regarding the type of networks that emerge in equilibrium. We now derive further implications of these findings on welfare and inequality.

\section{Income (Re)Distribution}

In this section, we first describe how the impact of free riding in networks depends on the initial income distribution. After characterizing the efficient solution, we derive policy implications on how a planner can redistribute income to increase welfare when agents can change their linking strategies as a result of the intervention. Finally, we show how the efficient solution can be implemented by using personalized prices.

\subsection{Income Distribution and Inequality}

When the equilibrium of the local public good game is a complete core-periphery network with inactive players in the periphery, spillovers flow as when the public good is global (i.e., among contributors and from contributors to free riders). However, there can be multiple equilibria, whose welfare properties can differ a lot across players depending on their network position. All players in the core, for example, consume the same bundle, while one richer player in the periphery can consume more than them, and thus, also obtain a larger utility (Proposition \ref{threshold}.\textit{(i)} and \textit{(ii)}).

For players in the periphery, it is key how many links they can afford. Each link frees additional resources the player can then spend on consuming more of both goods. Therefore, periphery players with more links benefit more from the network than those with less links: as richer players have more links, inequality increases.

By focusing on large societies, the next proposition can abstract from the welfare of core players, who represent an infinitesimal proportion of the population.
\begin{proposition}\label{inequality}
If Assumptions \ref{normality} and \ref{homogeneous} hold, for any sociable Nash equilibrium, there exists $\underline{w}$ such that\\
(i) if $\max_{i\in N}{w_i}-\min_{i\in N}{w_i}\leq \underline{w}$, the network reduces inequality in utility among any two players for a proportion of players that converges to 1 as $n\rightarrow\infty$;\\
(ii) otherwise, the network increases inequality in net social income among any two players for a proportion of players that converges to 1 as $n\rightarrow\infty$.\\
Furthermore, $\underline{w}$ is inversely related with $k$.
\end{proposition}
This result shows how the initial wealth distribution shapes inequality in payoffs if free riding in networks is costly. When societies are sufficiently homogeneous, networks are conducive to equality, since all periphery players can afford the same links, thereby enjoying the same spillovers. As utility is concave, richer players gain less marginal utility than poorer ones from these spillovers, and the difference in utility between any two periphery players is lower than absent the network. If to the contrary, the society is sufficiently unequal to begin with, then some players can afford more links, thereby free riding more. As a result, networks will only exacerbate the initial level of wealth inequality in the sense that the difference in net social income between periphery players is larger than the initial difference in wealth. Moreover, as in large societies the proportion of periphery players converges to 1, Proposition \ref{inequality} follows. 
This result hinges on the presence of a budget constraint and heterogeneity in income. When the budget constraint is not relevant, all players have the same free riding opportunities. In this case, as shown in the example in Figure \ref{comparison_aej}, the network always increases inequality when players have different valuations of the public good (\citealp{KM}, Proposition 2).

Finally, as $k$ decreases, the network reduces inequality even for a larger initial difference in wealth levels. The main reason for this is that net social income is increasing as $k$ decreases, and so poorer players also can take advantage of more free-riding opportunities in the network. This result highlights how lowering communication and transportation costs can increase welfare by allowing more people to take advantage of free riding opportunities.

\subsection{Efficiency}

We now characterize the efficient solution.
\begin{proposition}\label{efficiency}
If Assumption \ref{normality} holds, the efficient network is either empty or a star in which only the hub is active. A player is isolated if her valuation of the public good is too low.
\end{proposition}
The social planner solves two trade-offs: \textit{(i)} between consumption and links, since each additional link increases spillovers, but linking is costly; and \textit{(ii)} between private and public good consumption of the different players.

It is then quite intuitive that a star with only the hub contributing to the public good is the optimal solution as it minimizes linking costs while maximizing spillovers. 

While there is only one player in the core of the optimal solution, the model can accommodate for less stark results. If, for example, there were capacity constraints in links, the optimal solution would feature more than one player in the core. The reasoning on how to derive the optimal solution is the same.

It depends on preferences and total income whether and which players are isolated from the star. Indeed, excluding a player from the star is more beneficial the less she values the public good. Moreover, the number of links that the planner can afford depends on the total income available in the economy.

If all players have identical preferences and face identical prices, their marginal rate of substitution between public and private good consumption has to be equal. This implies that their ratios of marginal utilities per income are equal as well.
\begin{corollary}\label{eff_homo}
If Assumptions \ref{normality} and \ref{homogeneous} hold, in the efficient solution, their consumption of the private and public good are identical.
\end{corollary}

\subsection{Income Redistribution}

As decentralized equilibria are not efficient, we now ask which policies can the social planner introduce to increase welfare. We know from classical results that small transfers of income among contributors are neutral \citep{BBV}. However, this is true only under very specific assumptions when we consider local public goods \citep{nizar}. In our context, with endogenous networks, a transfer scheme is neutral not only when the set of contributors does not change, but also the transfer is confined to the largest contributors in the network and their provision does not change much, so that links are not affected. 

We next show how a planner can redistribute income to increase welfare. Income redistribution is a budget balanced transfers scheme $t = (t_1, ..., t_n)$ such that $\sum_{i \in N} t_i = 0$. Players then choose their optimal consumption and links according to the utility maximization problem \eqref{max} given $w+t$.

\begin{proposition}\label{improving_welfare}
Suppose a sociable equilibrium $(x^\ast,y^\ast,g^\ast)$ exists with non-empty $g^\ast$ and $\gamma^{\prime}_i\in(0,1]$ for all $i\in N$. Then, there is an income redistribution $t$ that yields an equilibrium with a star network $g^{star}$ in which both public good consumption and welfare are higher than in $(x^\ast,y^\ast,g^\ast)$.
\end{proposition}
In the proposition, we show that redistributing income towards the player with most links, eventually increases both public good consumption and welfare. This holds both for small transfers which do not change the equilibrium network, and for larger ones, which eventually induce a star as an equilibrium network. Notably, Proposition \ref{improving_welfare} allows for heterogeneous preferences, as long as the public good is strictly normal, which is a slightly stronger assumption than Assumption \ref{normality}.

This result builds on our characterization of equilibrium networks. Indeed, since sociable equilibrium networks are core-periphery graphs, spillovers among core players flow as in a complete network. Hence, a similar neutrality result as \cite{BBV}'s applies to this part of the network. In other words, consider the following thought experiment. First, fix the network; given this graph, we can transfer some resources from other core players to the player with the most links---who is also the largest contributor---in a way that each player's net social income is unchanged. Hence, their demand of the public good is unchanged. While the total provision of public good is then constant, the unique contributor in the core is the player who received the most links. As a result, periphery players' access to the public good increases, as they might not have been linked to all core players.

However, it is possible to further increase welfare and income by transferring resources to the player with most links until all the other core players are inactive and it is no longer profitable to link to them. A social planner can then collect the resources saved from deleting these links and use them to increase welfare. In particular, as $\gamma^{\prime}_i>0$ for all $i\in N$, part of these resources will be assigned to the largest contributor to further increase everyone's public good consumption.

Yet, the transfers we propose do not necessarily maximize welfare, for two reasons. First, when the planner transfers resources to the largest contributor, her net social income increases, but that of the player whose resources are being taken away decreases. So the optimal transfers need to trade-off the increases of public good provision of the hub with the decrease of private consumption of periphery players.

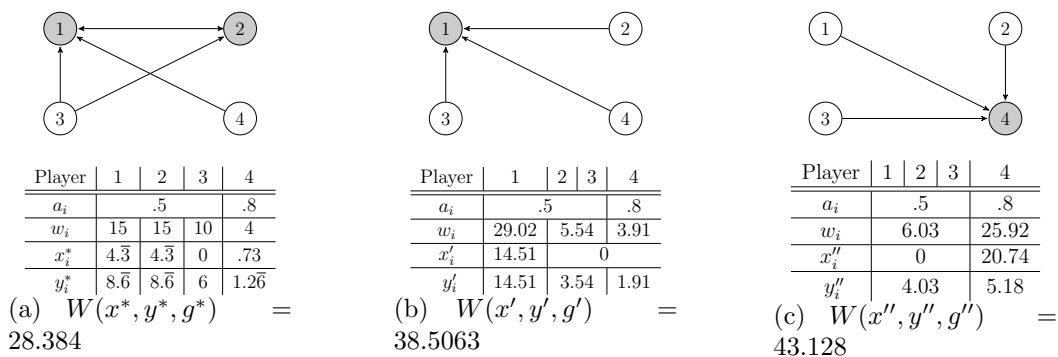
\begin{figure}[htbp!]
\begin{center}
\begin{minipage}{.33\linewidth}
\centering
\resizebox{3cm}{!}{
	\begin{tikzpicture}[-,>=stealth',shorten >=1pt,auto,node distance=2cm and 2cm,
          main node/.style={circle,draw}]

 \node[main node,fill=black!20] (1) {1};
 \node[main node,fill=black!20] (2) [right of=1, xshift=20mm] {2}; 
 \node[main node] (3) [below of=1] {3};
 \node[main node] (4) [below of=2] {4};

 \path[every node/.style={font=\sffamily\small}]
  (2) edge [<->] node [right] {} (1)
  (3) edge [->] node [right] {} (1)
  (3) edge [->] node [right] {} (2)
  (4) edge [->] node [right] {} (1);
\end{tikzpicture}
	}
\end{minipage}
\begin{minipage}{.33\linewidth}
\centering
\resizebox{3cm}{!}{
	\begin{tikzpicture}[-,>=stealth',shorten >=1pt,auto,node distance=2cm and 2cm,
          main node/.style={circle,draw}]

 \node[main node,fill=black!20] (1) {1};
 \node[main node] (2) [right of=1, xshift=20mm] {2}; 
 \node[main node] (3) [below of=1] {3};
 \node[main node] (4) [below of=2] {4};

 \path[every node/.style={font=\sffamily\small}]
  (2) edge [->] node [right] {} (1)
  (3) edge [->] node [right] {} (1)
  (4) edge [->] node [right] {} (1);
\end{tikzpicture}
	}
\end{minipage}
\begin{minipage}{.32\linewidth}
\centering
\resizebox{3cm}{!}{
	\begin{tikzpicture}[-,>=stealth',shorten >=1pt,auto,node distance=2cm and 2cm,
          main node/.style={circle,draw}]

 \node[main node] (1) {1};
 \node[main node] (2) [right of=1, xshift=20mm] {2}; 
 \node[main node] (3) [below of=1] {3};
 \node[main node,fill=black!20] (4) [below of=2] {4};

 \path[every node/.style={font=\sffamily\small}]
  (2) edge [->] node [right] {} (4)
  (3) edge [->] node [right] {} (4)
  (1) edge [->] node [right] {} (4);
\end{tikzpicture}
	}
\end{minipage}

\vspace{.3cm}

\begin{minipage}{.33\linewidth}
\centering
\subfigure[$
W(x^\ast,y^\ast,g^\ast)=
28.384$]{\label{1transf}
\resizebox{3.5cm}{!} {
\begin{tabular}{c|c|c|c|c}
 Player & 1 & 2 & 3 & 4 \\ 
\hline \hline $a_i$ & \multicolumn{3}{c|}{.5} & .8 \\
\hline $w_i$ & 15 & 15 & 10 & 4\\
\hline \rule{0pt}{12pt} 
$x_i^{\ast}$& $4.\overline{3}$ & $4.\overline{3}$ & $0$ & $.73$ \\
\hline \rule{0pt}{12pt}
$y_i^{\ast}$& $8.\overline{6}$ & $8.\overline{6}$ & $6$ & $1.2\overline{6}$
\end{tabular}
}
	}
\end{minipage}
\begin{minipage}{.33\linewidth}
\centering
\subfigure[$
W(x^\prime,y^\prime,g^\prime)=
38.5063$]{\label{2transf}
\resizebox{3.5cm}{!} {
\begin{tabular}{c|c|c|c|c}
 Player & 1 & 2 & 3 & 4 \\ 
\hline \hline $a_i$ & \multicolumn{3}{c|}{.5} & .8 \\
\hline $w_i$ & $29.02$ & \multicolumn{2}{c|}{$5.54$} & 3.91 \\
\hline $x_i^{\prime}$& $14.51$ & \multicolumn{3}{c}{0}\\
\hline \rule{0pt}{12pt}
$y_i^{\prime}$& $14.51$ & \multicolumn{2}{c|}{$3.54$} & $1.91$
\end{tabular}
}
	}
\end{minipage}
\begin{minipage}{.32\linewidth}
\centering
\subfigure[$
W(x^{\prime\prime},y^{\prime\prime},g^{\prime\prime})=
43.128$]{\label{3transf}
\resizebox{3.5cm}{!} {
\begin{tabular}{c|c|c|c|c}
 Player & 1 & 2 & 3 & 4 \\ 
\hline \hline $a_i$ & \multicolumn{3}{c|}{$.5$} & $.8$ \\
\hline $w_i$& \multicolumn{3}{c|}{$6.03$} & $25.92$ \\
\hline $x_i^{\prime\prime}$& \multicolumn{3}{c|}{0} & $20.74$ \\
\hline \rule{0pt}{12pt}
$y_i^{\prime\prime}$& \multicolumn{3}{c|}{$4.03$} & $5.18$
\end{tabular}
}
	}
\end{minipage}

\caption{Welfare improving vs. welfare maximizing transfers for $k=2$, $U_i(\overline{x}_i,y_i) ={\overline{x}}_i^{a_i} y_i^{1-a_i}$ and $p_i=1$ for all $i\in N$. Players in the core are shaded in gray.
}\label{ex_transfers}
\end{center}
\end{figure}

Second, the player who becomes the hub of the star applying the transfers described in Proposition \ref{improving_welfare} might not be the ideal recipient. To clarify the ideas, consider the example in Figure \ref{ex_transfers}. Panel (a) represents the initial allocation, while panel (b) the optimal allocation given that player 1 is the hub of the star. Panel (c) instead represents the welfare maximizing allocation, where resources are being transferred to player 4, who has the highest valuation for the public good. Indeed, player 4 is initially in the periphery because of a limited initial budget.\footnote{A similar situation can arise when players pay different prices for the private good.} This subtle difference is very important, as it implies that repeatedly applying the redistribution policies characterized by \cite{nizar} 
to the initial equilibrium would not lead to the welfare-maximizing outcome.

The following proposition addresses these concerns. We define welfare-maximizing transfers as those yielding the Nash equilibrium with the maximal welfare among all Nash equilibria arising from all budget-balanced transfers.
We now characterize the welfare-maximizing transfer scheme taking into account network endogeneity.\footnote{
As shown in Proposition \ref{efficiency}, a star is the efficient solution. However, the welfare-maximizing transfer scheme is associated with a lower public good provision than in the efficient solution whenever $\gamma^{\prime}_i<1$, where $i$ is the hub of the star network.}
\begin{proposition}\label{2ndBest}
Suppose there exists a non-empty sociable equilibrium network $g^*$. Then, there is $\Gamma>0$ and welfare-maximizing transfers if $\gamma^{\prime}_i\in (0,\Gamma]$ for all $i\in N$. 
\end{proposition}
This result is obtained with few assumptions on players' preferences except for a condition on how steep each player's Engel curve is at the second best. Intuitively, this condition requires that the demand of periphery players cannot be too high when they are poor and needs to be relatively low if they are rich, as players who are in the periphery in the second best would provide less public good if they were the hub. Hence, each player's Engel curve cannot be too concave. 

To show that this requirement is not too strict, we show in the next corollary that the welfare-maximizing transfer scheme can always be Nash implemented if players have identical and linear Engel curves.
\begin{corollary}\label{linear}
Suppose players have homogeneous preferences with linear Engel curves such that $\gamma^{\prime}_i\in(0,1]$ for all $i\in N$, and a sociable equilibrium $(x^\ast,y^\ast,g^\ast)$ exists with non-empty $g^\ast$. Then, the second best can always be implemented as a Nash equilibrium.
\end{corollary}
Finally, by Corollary \ref{ex-bbv}, the welfare-maximizing transfer scheme induces a unique equilibrium if the linking cost is sufficiently low.

\subsection{Personalized Prices}

We have shown so far that income redistribution can go a long way in increasing welfare. However, the resulting allocation is not efficient. Indeed, the planner cannot dictate how players use the resources allocated to them, which results in a much higher private consumption of the private good with respect to the efficient one, as derived in Proposition \ref{efficiency}.

We now go one step further and show that a personalized lump-sum tax scheme can instead achieve the efficient outcome. Indeed, the optimal policy here subsidizes the price of the public good for the player selected as hub in Proposition \ref{efficiency}. The personalized price is set equal to this player's marginal rate of substitution in the efficient solution. This induces her to choose the socially optimal public good provision and all other players to free ride on her provision. The lump-sum taxation scheme is designed to be budget balanced. 
\begin{proposition}\label{personalized prices}
Suppose $\gamma^{\prime}_i\in(0,1)$ for all $i\in N$ and a sociable equilibrium $(x^\ast,y^\ast,g^\ast)$ with non-empty $g^\ast$ exists. Then, the efficient solution can be implemented by fixing a personalized price of the public good $p_x \in (0,1)$ for the player selected as hub in Proposition \ref{efficiency}, financed by lump-sum taxes from all players.
\end{proposition}
Since only one player provides the public good, this situation is analogous to a direct government provision of a public good that crowds out private contributions. However, two additional features are relevant. First, crowding-out is incomplete, so that a public intervention can increase welfare by increasing public good provision. Most importantly, an additional increase in consumption can be achieved by inducing a star, thereby re-directing resources from linking to consumption. 


Proposition \ref{personalized prices} is relevant whenever the social planner is able to affect the price of the public good. In the context of online social platforms, this could take the form of subsidizing content of the ``super-star'' users who contribute a lot to the content of the platform. Reward schemes in line with this idea are indeed implemented for example by YouTube, whose main contributors receive awards and are sponsored to participate in various promotional activities.

\section{Robustness and Extensions}\label{extensions}

Our characterization is robust to the introduction of selfish motives in the private provision of public goods due to warm glow giving, e.g., because of the private benefits of becoming an influencer on Twitter. Our characterization is robust also to more general linking technologies, an imperfect degree of substitutability between neighbors' public good provision, and if players benefit from global contributions to the public good, for example, because they can see the posts of others they are not linked to using hashtags on Twitter. Additionally, the characterization of strict Nash equilibria is robust to the introduction of some complementarity in players' efforts, to heterogeneity in the costs of forming a connection and to indirect spillovers, e.g., re-tweeting posts.

To capture situations where a link between two players implies that both access each others' public good provision, we also extend the model to \textit{two-way flow of spillovers}. Some examples are the acquisition and exchange of information about new products and technologies.

In \cite{gg} and \cite{KM}, the demand for the public good is exogenous. In those models, when players are homogeneous, strict Nash networks are complete core-periphery networks, while, when players are heterogeneous, the prediction depends on how the gains from a connection change across players. Furthermore, the law of the few holds.

We extend their models to a more general class of preferences and heterogeneous budgets. While most of our results carry over to the two-sided model, the key difference from the model presented above is that core players benefit ``for free'' from the public good provided by those linking to them. In this case, 
strict Nash equilibria need not be complete core-periphery networks even when players are homogeneous. Additionally, players in the core have the same consumption only if they have the same neighbors, and a richer player in the periphery is worse off than all core players.


When spillovers flow two-way, a stronger version of the law of the few of Proposition \ref{lotf} holds: the proportion of players significantly contributing to the public good (even in the periphery) goes to zero as the population size grows to infinity. Importantly, we highlight the necessary condition for this result to hold, i.e., that the slope of the contributors' Engel curve for the public good is strictly below one. In that case, contributors' provision strictly decreases with spillovers from neighbors. As a result, the amount of spillovers received by contributors cannot be infinitely large.

All of these results are formally derived in the Online Appendix.

\section{Conclusions}\label{conclusions}

In this paper, we introduce a model in which players decide how to allocate their budget between links, a local public good and a private good. A player links to another in order to free ride on her public good provision. Under standard assumptions, Nash equilibrium networks are core-periphery graphs, a prediction that is robust to several extensions. Importantly, since the demand for the public good is endogenous, poorer players can attain more central positions in the network or consume more public good. Hence, the relationship between wealth, provision, consumption and, therefore, utility is not as straightforward as when the public good is global.


We have shown that the type of equilibrium that emerges has an impact on welfare and that income redistribution can alleviate inefficiencies. Knowledge of players' preferences is key in the design of optimal transfers. Yet, as stating a high taste for the public good would translate into receiving more resources, future research should investigate how to incentivize people to truthfully state the taste for the public good of their neighbors and their own. Indeed, players observe better than the social planner their neighbors' preferences. Recent literature has shown that there is scope for eliciting information from agents in a fixed network \citep{bloch2021friend,baumann2018self}. It would be interesting, however, to develop similar insights when the mechanism needs to take into account also link formation.


\appendix

\setcounter{equation}{0}
\setcounter{proposition}{0}
\newcommand\appendixlabel{A}
\renewcommand{\theequation}{A-\arabic{equation}}

\section*{Appendix: Proofs}\label{Proofs}

\noindent \textbf{Proof of Lemma \ref{contribution}.} Since $\gamma_i^{\prime}\in[0,1]$ and $x_i=\gamma_i(w_i-\eta_i k+x_{-i})-x_{-i}$ for all $i \in N$, we have that $dx_i/dx_{-i}=\gamma_i^{\prime}(w_i-\eta_i k+x_{-i})-1<0.$
\hfill $\blacksquare$

\bigskip

\noindent \textbf{Proof of Lemma \ref{almost_core}.} Suppose that $g^{\ast}_{ji}=0$. Since $x^{\ast}_j=\gamma_j(\overline{w}_j(g^{\ast}))-x^{\ast}_{-j} >k$, by Assumption \ref{normality}, if $j$ links to $i$, $j$'s public good consumption is higher since $\gamma_j(\overline{w}_j(g^{\ast})-k+x^{\ast}_i)>\gamma_j(\overline{w}_j(g^{\ast}))$ given that $x^{\ast}_i>k$ and $\gamma_j^{\prime}\geq 0$. Suppose that this is not true. Then, a contradiction arises as either this is no equilibrium, or $x_i^* < k$, or both. Since $\gamma_j^{\prime}\leq 1,$ also $j$'s consumption of the private good increases if $j$ links to $i$. This contradicts $g^{\ast}_{ji}=0$. This concludes the proof of Lemma \ref{almost_core}. \hfill $\blacksquare$

\bigskip

\noindent \textbf{Proof of Proposition \ref{characterization}.} If $k>\overline{k}=max_{i\in N}\{x_i^I\}$, no link is profitable, so the unique equilibrium is the empty network with all players $i\in N$ consuming $(x^I,y^I)$.

\noindent If $k\leq \overline{k}$, suppose that there is no player $j\in N$ such that $x_j^{\ast}=k$. 
Then, the next lemma shows that there is a unique non-empty component.
\begin{lemma}\label{nonempty}
If there is $i\in N$ such that $x_i^{\ast} > k$, then $g^{\ast}$ is either empty or has a unique component. If there is $j \neq i$ such that $x_j^I > k$, 
then $g^{\ast}$ is non-empty.
\end{lemma}
\textbf{Proof of Lemma \ref{nonempty}.} Consider an equilibrium $(x^{\ast},y^{\ast},g^{\ast})$. 
If it is not profitable for any player to link to $i$, the network is empty. If the network is non-empty and there is more than one component, there is $j \neq i$ with $\bar{g}_{ji}^*=0,$ and $g_{jz}^* = 1$ such that $i \neq z$, while $\bar{g}_{iz}^* = 0$. Without loss of generality $x^{\ast}_z \leq x_i^{\ast}$. If $x^{\ast}_z < x_i^{\ast}$, it is profitable for $j$ to sever the link to $z$ and link to $i$ instead. If $x^{\ast}_z = x_i^{\ast}$, it is profitable for $z$ to link to $i$ since $z$ would access the same amount of public good at a lower cost. This proves the first part of Lemma \ref{nonempty}, while the second follows from Lemma \ref{almost_core}, concluding the proof of Lemma \ref{nonempty}. \hfill $\blacksquare$

\noindent Since $g^{\ast}$ is non-empty and $x_j^{\ast} \neq k$ for all $j\in N$, there is $i$ such that $x_i^{\ast} > k$. By Lemma \ref{almost_core}, $\overline{g}^\ast_{ij}=1,$ if $x_j^\ast \geq x_i^\ast > k.$ Hence, a core exists such that $i \in \mathcal{C}$ if and only if $x_i^{\ast} > k$. For any player $i \notin \mathcal{C}$, $x_i^{\ast} < k$; these players are either isolated or link to players in the core. This concludes the proof of the first part of Proposition \ref{characterization}.

\noindent Hence, any non-empty equilibrium network is a core-periphery graph, in which $i \in \mathcal{C}$ if and only if $x_i^\ast >k.$ Suppose that in a strict equilibrium $g^{\ast}$ is not a nested split graph. Then, there are players $i,j \in \mathcal{C}$ and $p,z\in\mathcal{P}$ such that $g_{zj}^{\ast}=g_{pi}^{\ast}=0$ and $g_{pj}^{\ast}=g_{zi}^{\ast}=1$. Since the equilibrium is strict, assume without loss of generality $x_j^{\ast}> x_i^{\ast}$. Then, $z$ can profitably sever the link with $i$ and link to $j$ instead, a contradiction. This concludes the proof of Proposition \ref{characterization}. \hfill $\blacksquare$

\bigskip


\noindent \textbf{Proof of Proposition \ref{existence}.} By Lemma \ref{contribution}, in equilibrium a player $i\in N$ such that $x_i^I< k$ never produces more than $k$, and hence never receives links.

\noindent We now prove that an equilibrium exists whose associated network has a core-periphery structure first for any $\gamma_i^\prime \in (0,1]$. Denote by $(x(g^{D}),y(g^{D}),g^{D})$ an allocation where $(x(g^{D}),y(g^{D}))$ are optimal given $g^{D}$ for all $i\in N$ and $g^{D}$ is a network such that $g_{ij}=g_{ji}=1$ for all $i,j\in D$, while $g_{zi}=1$ for $i\in D$ and $z\in \mathcal{P}$ if and only if $U_z(x(g^{D}),y(g^{D}),g^{D})\geq U_z((x_z^{\prime},x_{-z}(g^{D})),(y_z^{\prime},y_{-z}(g^{D})),g^{D}-g_{zi})$ for any feasible $(x_z^{\prime},y_z^{\prime})$, where $g^{D}-g_{zi}$ is a network resulting from deleting the link $g_{zi}$ from $g^{D}$. In other words, $D$ is the set of core players, while the other players only establish the links they find profitable to players in $D$. So, all players are best responding in terms of private and public good consumption.

\noindent For each player $i\in D$, public good consumption $\overline{x}_i(g^D)$ is given by
\begin{eqnarray}\label{ind_provision}
\overline{x}_i(g^D)=\gamma_i(w_i- \eta_i(g^D) k+x_{-i}(g^D))
\end{eqnarray}
where we denote $\phi_i=\gamma_i^{-1}$ and by $\overline{x}^D$ the quantity of public good consumed by each player in the core. Let $|D|=\eta_i(g^D)+1$ for $i\in D$ be the size of the core. Then, $\phi_i(\overline{x}^D)=w_i-(|D|-1) k+x_{-i}(g^D)$. Summing this equation for all $i\in D$, we get
\begin{eqnarray}\label{total_provision}
\sum_{i\in D} \phi_i(\overline{x}^D)+(1-|D|) \overline{x}^D=\sum_{i\in D} w_i- |D| (|D|-1) k.
\end{eqnarray}
Denote $F(|D|,\overline{x}^D)=\sum_{i\in D} \phi_i(\overline{x}^D)+(1-|D|) \overline{x}^D$. First consider $\gamma_i^{\prime}>0$ for all $i\in N$. Given a network $g^D$, for players in $D$, Theorem 3 in \cite{BBV} applies: $F(|D|,\overline{x}^D)$ is strictly increasing in $\overline{x}^D$, so that there exists a unique equilibrium in private and public good consumption for players in $D$. Consider then $\mathcal{D}=\{D\subset N | x_i(g^D)\geq k$ for all $i\in D\}$, where $\mathcal{D}\neq \emptyset$ as there exists $i\in N$ such that $x_i^I\geq k$ since $k\leq\overline{k}$. However, the smaller is $k$, the larger $\max |D|$ for $D\in \mathcal{D}$ can be. Consider then one such set $D$. By Lemma \ref{contribution}, if $\gamma^{\prime}_p<\overline{\gamma}$, $x_p(g^D)<k$ for all $p\in \mathcal{P}$. By construction, $x_i(g^D)\geq k$ and $g_{ij}=1$ for all $i,j\in D$, as required by Lemma \ref{almost_core}. Hence, $(x(g^{D}),y(g^{D}),g^{D})$ is an equilibrium.

\noindent If $\gamma_i^{\prime}=0$ for some $i\in N$, then player $i$ has a fixed demand for the public good. Thus, $i$ spends her budget on linking to the biggest contributors, and if needed, with the remaining budget provides public good. Hence, $i$ is in the periphery if players in the core satisfy her demand, or in the core otherwise. In either case, this is irrelevant for existence. If $\gamma^{\prime}_i=0$ for all $i\in N$, then a star with $i=\arg\max_{i\in N} x_i^I$ as the hub is trivially an equilibrium. This concludes the proof of Proposition \ref{existence}. \hfill $\blacksquare$

\bigskip

\noindent \textbf{Proof of Corollary \ref{ex-bbv}.} If $k=0$, every player is connected to any player with a positive contribution; any other link is payoff irrelevant. Hence, the unique sociable equilibrium is a complete network, and by Theorem 3 in \cite{BBV}, a unique set of contributors exists, which we denote by $D$. 

\noindent Given this set of contributors $D$, equations \eqref{ind_provision} and \eqref{total_provision} indicate that both individual and total provision decrease continuously as $k$ increases if $\gamma_i^\prime \in (0,1)$ for all $i\in N$. 
Hence, there exists $\tilde{k}\geq 0$ such that for any $k\leq \tilde{k}$, $x_i^{D}\geq k$ for all $i\in D$. Then, for any $k\leq \tilde{k}$, there exists an equilibrium in which the set of contributors is $D$, each producing at least $k$, these contributors form a clique, and periphery players are inactive and link, starting with the highest producing player, to all or some of the contributors depending on their utility function and wealth. The resulting network is a core-periphery network, and it is easy to see that it is an equilibrium and it is unique. This concludes the proof of Corollary \ref{ex-bbv}. \hfill $\blacksquare$


\bigskip

\noindent \textbf{Proof of Proposition \ref{lotf}.} Consider any sociable Nash equilibrium $(x^{\ast},y^{\ast},g^{\ast})$. If $g^{\ast}$ is empty, the statement follows trivially. Otherwise, for all $z \in \mathcal{C}$, $x_z^{\ast}>k$ and, by Lemma \ref{almost_core}, they are all linked among themselves. Suppose that $\lim_{n\rightarrow\infty}|\mathcal{C}|/n>0$. Each $i\in \mathcal{C}$ spends $k|\mathcal{C}|$ on links, which would converges to infinity as $n\rightarrow\infty$. Since $w_i<\omega$ for all $i\in N$, this is a contradiction, thus proving Proposition \ref{lotf}. \hfill {$\blacksquare $}

\bigskip

\noindent \textbf{Proof of Proposition \ref{threshold}.} Order the players so that $w_1 \geq w_2 \geq ... \geq w_n$. Since $\gamma_i = \gamma$ for all $i \in N$, $x_1^I \geq x_2^I\geq ... \geq x_n^I$. We prove first the following lemma, that we use to show existence.

\begin{lemma}\label{inthecore}
Suppose $\gamma_i=\gamma$ for all $i\in N$. Consider an equilibrium $(x(g^D),y(g^D)g^{D})$ such that $x_i(g^{D})\geq k$ for all $i\in D\cup \{z\}$ such that $w_i\geq w_z$ for all $i\in D$ and $z\in\mathcal{P}(g^D)$. Then, $x_i(g^{D\cup \{z\}})\geq k$ for all $i\in D\cup \{z\}$.
\end{lemma}
\noindent \textbf{Proof of Lemma \ref{inthecore}.} Since $x_z(g^{D})>0$, $ x_z(g^{D})=\gamma (w_z-|D| k+x_{-z}(g^{D}))-x_{-z}(g^{D})$. Similarly, $ x_z(g^{D\cup \{z\}})=\gamma (w_z-|D| k+x_{-z}(g^{D\cup \{z\}}))-x_{-z}(g^{D\cup \{z\}})$. Then, $x_z(g^{D\cup \{z\}})\geq x_z(g^{D})$ if and only if $x_{-z}(g^{D\cup \{z\}})\leq x_{-z}(g^{D})$.

\noindent Suppose now $x_z(g^{D\cup \{z\}})<k$. This would imply that $x_{-z} (g^{D\cup \{z\}})> x_{-z} (g^{D})$. Since $x_i (g^{D})=\gamma(w_i-(|D|-1) k+x_{-i} (g^{D}))-x_{-i} (g^{D})$ for any $i \in D,$ and $x_z(g^{D\cup \{z\}})< k$, $w_i-|D|k+x_z(g^{D})\leq w_i-(|D|-1)k$, by Lemma \ref{contribution}, $x_{-z} (g^{D\cup \{z\}})\leq x_{-z} (g^{D})$, a contradiction. Hence, $x_z(g^{D\cup \{z\}})\geq x_z(g^{D})\geq k$.

\noindent Finally, since $\overline{x}_i (g^{D\cup \{z\}})=\overline{x}_z (g^{D\cup \{z\}})$ for $i\in D$, $w_i-|D|k+x_{-i} (g^{D\cup \{z\}})=w_z-|D|k+x_{-z} (g^{D\cup \{z\}})$, since $w_i \geq w_z$, $x_{i} (g^{D\cup \{z\}})\geq x_{z} (g^{D\cup \{z\}})\geq k$. This concludes the proof of Lemma \ref{inthecore}. \hfill $\blacksquare$

\noindent Consider first the empty network. If no player wants to link to player 1, this is an equilibrium. Otherwise, $x^I_1\geq k$ and let all players willing to link to $1$ do that. The resulting network is $g^1$.
If $x_i^*(g^{1})<k$ for all $i\in \mathcal{P}(g^{1}),$ then $g^{1}$ is an equilibrium. Otherwise, $x_2^*(g^{1})\geq k$. By Lemma \ref{inthecore}, $1$ and $2$ produce more than $k$ in $g^{\{1,2\}}$. If $x_i^*(g^{\{1,2\}})<k$ for each $i\in \mathcal{P}(g^{\{1,2\}})$, $g^{\{1,2\}}$ is an equilibrium. Otherwise, note that: first, if someone is producing more than $k$ and best-replying, she must be already linked to all players producing more than $k$ (Lemma \ref{almost_core}); second, a player's contribution is increasing in wealth (Lemma \ref{contribution}). Hence, if $g^{\{1,2\}}$ is no equilibrium, $x_3^*(g^{\{1,2\}})\geq k$. We can then move $3$ to the core, and by Lemma \ref{inthecore}, $x_c^*(g^{\{1,2,3\}})\geq k$ for $c=\{1,2,3\}$.
This argument is recursive until there is a network in which all players in the core produce more than $k$ and all players in the periphery (if any) produce less than $k$. By construction, this constitutes a Nash equilibrium, and proves existence.

\noindent Note that, since Lemma \ref{inthecore} holds for $\gamma^{\prime} \in \{0,1\}$, so does the recursive argument. 
Suppose that $\gamma_i^\prime = 0$ for all $i \in N$. Then, either the network is empty or by the same procedure as before a network with a core exists and is a Nash equilibrium (since any player's demand is this case is fixed). Suppose instead that $\gamma_i^\prime = 1$ for all $i \in N$. Then, any $i$'s provision of the public good only depends 
on her income $w_i$, but not on spillovers. Again an equilibrium exists by the previous argument.

\noindent Ad \textit{(i)} By Lemma \ref{almost_core} and the fact that $i,j \in \mathcal{C},$ both $\overline{x}_i^* = \overline{x}_j^*$ and $\eta_i = \eta_j$. Since $\overline{x}_i^* = \gamma(w_i - \eta_i k + \overline{x}_{-i}^*)$ and $\gamma_i = \gamma$ for all $i \in N,$ $\overline{w}_i \equiv \tilde{w}$ for all $i \in \mathcal{C}.$ Since $x_i^* = \gamma (\tilde{w}) - \overline{x}_{-i}^*,$ and $y_i^* = w_i - \eta_i k - x_i^* = \tilde{w} - \gamma(\tilde{w})$, for $i,j \in \mathcal{C},$ $y_i^* = y_j^*$ which shows the first part of \textit{(i)}. Finally, $x_i^* = \gamma(w_i - \eta_i k + \overline{x}_{-(i \cup j)}^* + x_j^*) - \overline{x}_{-(i \cup j)}^* - x_j^*$, and $x_j^* = \gamma(w_j - \eta_j k + \overline{x}_{-(i \cup j)}^* + x_i^*) - \overline{x}_{-(i \cup j)}^* - x_i^*.$ Solving for $\overline{x}_{-(i \cup j)}^*$ yields $\gamma(w_i - \eta_i k + \overline{x}_{-(i \cup j)}^* + x_j^*) - x_i^* - x_j^* = \gamma(w_j - \eta_j k + \overline{x}_{-(i \cup j)}^* + x_i^*) - x_j^* - x_i^*,$ or $w_i + x_j^* = w_j + x_i^*.$ So $w_j > w_i,$ if and only if $x_j^* > x_i^*$.\\
\noindent Ad \textit{(ii)} If $x_j^* = 0$, $\overline{x}_j^* = \overline{x}_i^*$ since $g_{ji}^* = 1$ for all $i \in \mathcal{C}.$ If $x_j^* > 0$, then $\overline{x}_j^* > \overline{x}_i^*.$\\
\noindent Ad \textit{(iii)} Since $\overline{g}^*_{ij} = 0$ for $i,j \in \mathcal{P}$, $N_i = N_j$ implies that $\eta_i = \eta_j$ and $\overline{x}_{-i}^* = \overline{x}_{-j}^*$. Given $\gamma_i = \gamma$ for all $i \in N$, $x_i^* = \gamma(w_i - \eta_i k + \overline{x}_{-i}^*) \geq \gamma(w_j - \eta_j k + \overline{x}_{-j}^*) = x_j^*$, $\overline{x}_i^* \geq \overline{x}_j^*$ and $y_i^* \geq y_j^*$ if, and only if, $w_i \geq w_j$.\\
\noindent Ad \textit{(iv)} Suppose that $i,j \in \mathcal{P}$ and $w_i \geq w_j.$ Then, given $\gamma_i = \gamma$ for all $i \in N,$ $\overline{x}_i^* \geq \overline{x}_j^*$. Then, clearly $N_i \geq N_j$.\\
\noindent Ad \textit{(v)} Define the set of equilibria as $E(x^*, y^*, g^*)$, and suppose that for all $g^* \in E(\cdot)$, $i \in \mathcal{P}.$ Collect all such players $i$ in set $PE$ and then choose $w^\prime = \max_{i \in PE} w_i.$ \hfill $\blacksquare$



\bigskip

\noindent \textbf{Proof of Corollary \ref{richer}.} Define the total public good contribution of players in the core of a core-periphery network $g$ as $x^{\mathcal{C}(g)}$, where $x^{\mathcal{C}(g)}$ solves
\begin{eqnarray}\label{total_provision_h}
|\mathcal{C}(g)| \phi_i(\overline{x}^{\mathcal{C}(g)})+(1-|\mathcal{C}(g)|) \overline{x}^{\mathcal{C}(g)}=\sum_{i\in \mathcal{C}(g)} w_i- |\mathcal{C}(g)| (|\mathcal{C}(g)|-1) k,
\end{eqnarray}
with $\phi_i=\gamma_i^{-1}$. If $\gamma^{\prime}\in(0,1]$, the LHS of \eqref{total_provision_h} is strictly increasing in $x^{\mathcal{C}(g)}$ (Corollary \ref{ex-bbv}), while the RHS is strictly increasing in the total wealth of core players and strictly decreasing in $k$ (as $\partial RHS/\partial k=-2n$, when there is an additional player, $2n$ more links have to be established). By Proposition \ref{threshold}.\textit{(iv)}, there exist $\omega>0$ such that if $\max_{i\in N}{w_i}-\min_{i\in N}{w_i}<\omega$, everyone is linked to all players in the core. So, welfare in $g^h$ is higher than in $g^{\prime}$ if and only if $x^{\mathcal{C}(g^h)}>x^{\mathcal{C}(g^\prime)}$. Suppose that $|\mathcal{C}(g^h)|=|\mathcal{C}(g^{\prime})|$. Then, trivially $x^{\mathcal{C}(g^h)}>x^{\mathcal{C}(g^\prime)}$ as $g^h$ has richer players in the core. Suppose now that $|\mathcal{C}(g^h)|<|\mathcal{C}(g^{\prime})|$. If $\sum_{i\in \mathcal{C}(g^h)} w_i>\sum_{i\in \mathcal{C}(g^\prime)} w_i$, then $x^{\mathcal{C}(g^h)}>x^{\mathcal{C}(g^\prime)}$. Otherwise, since the LHS of \eqref{total_provision_h} is strictly decreasing in $k$, there is $K$ such that $x^{\mathcal{C}(g^h)}>x^{\mathcal{C}(g^\prime)}$. 
It follows from Proposition \ref{threshold}.\textit{(i)} and Lemma \ref{contribution} that $|\mathcal{C}(g^h)|>|\mathcal{C}(g^{\prime})|$ is not part of an equilibrium. This concludes the proof of Corollary \ref{richer}. \hfill $\blacksquare$

\bigskip

\noindent \textbf{Proof of Proposition \ref{inequality}.} Consider any sociable Nash equilibrium $(x^{\ast},y^{\ast},g^{\ast})$. By part \textit{(iii)} and \textit{(iv)} of Proposition \ref{threshold}, there exists $\underline{w}$ such that if $\max_{i\in N}{w_i}-\min_{i\in N}{w_i}\leq \underline{w}$, $N_i(g^*) = N_j(g^*)$ for any $i,j \in \mathcal{P}$; in this case, $\overline{w}_i - \overline{w}_j = w_i - w_j$. Furthermore, in any non-empty network, $\overline{w}_i(g^\ast)>w_i$ for any $i\in \mathcal{P}$. Given the concavity of the utility function, for $i,j\in\mathcal{P}$ such that $w_i\geq w_j$, $U(\overline{x}_i^\ast,y_i^\ast,g^\ast) - U(\overline{x}_j^\ast,y_j^\ast,g^\ast) \leq U(x_i^I,y_i^I,g^\emptyset) - U(x_j^I,y_j^I,g^\emptyset)$, where $g^\emptyset$ is the empty network, with equality if and only if $w_i=w_j$.
If instead $\max_{i\in N}{w_i}-\min_{i\in N}{w_i}> \underline{w}$, there exist $i,j \in \mathcal{P}$ such that $N_i(g^*) > N_j(g^*)$; hence, for any pair of such players, $\overline{w}_i - \overline{w}_j > w_i - w_j$. To complete this part of the proof, note that by Proposition \ref{lotf}, the proportion of players in the periphery converges to 1 as $n\rightarrow\infty$. 

Finally, given $\tilde{k}$, suppose that $\max_{i\in N}{w_i}-\min_{i\in N}{w_i}> \underline{w}(\tilde{k})$ and $N_i(g^*) \subset N_j(g^*)$ for $i,j \in \mathcal{P}$, while for all $l \neq i$, $N_j(g^*) = N_l(g^*)$; i.e., there is $z \in \mathcal{C}$ with $g_{iz}^* =0$, while $g_{jz}^* = 1$ for all $j \neq i$. Now decrease linking cost to $\underline{k} <\tilde{k}$ until $g_{iz}^* =1$, i.e., $i$ is just indifferent to link to $z$ or not, but in a sociable Nash equilibrium she does. Then, $N_i(g^*) = N_j(g^*)$ for all $i,j \in \mathcal{P}$, and $\underline{w}(\underline{k}) \geq \max_{i\in N}{w_i}-\min_{i\in N}{w_i} > \underline{w}(\tilde{k})$. So while incomes did not change, social incomes did, and, due to the decrease in $k$, inequality is also reduced. Hence, we conclude that $\underline{w}$ is inversely related with $k$. This concludes the proof of Proposition \ref{inequality}. \hfill {$\blacksquare $}

\bigskip

\noindent \textbf{Proof of Proposition \ref{efficiency}.} If the efficient network is non-empty, it is a collection of stars in which each hub is the only active player, since this minimizes linking costs. If there are several stars, consider two hubs $i$ and $j$ in two of them. Then, $x_i, x_j > k,$ and transferring $i$'s public good provision to $j$, rewiring all in-links of $i$ to $j$, and linking $i$ to $j$ increase consumption for all players. Hence, if the network is non-empty, connected players in the efficient solution form a star with only one active player. Denote this network by $g^{star}$ and its hub by $h$. 

\noindent If all players belong to the star, the efficient (first-best) solution $(x^{fb},y^{fb},g^{star})$ solves
\begin{eqnarray}\label{maxeffcons}
\max_{x_h, y_1, ..., y_n} && \sum_{i\in N} U_i(x_h,y_i)
\text{ s.t. } x_h + \sum_{i \in N} y_i \leq \sum_{i\in N} w_i -(n-1) k, 
\end{eqnarray}
leading to the following FOCs:
\begin{eqnarray*}
\sum_{i \in g^{star}} \frac{\partial U_i (x_h^{fb},y_i^{fb}) }{ \partial x}=\lambda, \\
\frac{\partial U_i (x_h^{fb},y_i^{fb}) }{ \partial y_i}=\lambda \text{ for all } i\in g^{star},\\
x_h^{fb}= \sum_{i \in g^{star}} w_i -(n-1)k - \sum_{i \in g^{star}} y_i^{fb}, 
\end{eqnarray*}
where $\lambda$ is the shadow value of the social planner's budget constraint.

\noindent Suppose next that $j$ is removed from the star. It is efficient to severe $g_{jh}$ if
\begin{eqnarray}\label{obf2}
\sum_{i\in g^{star} \setminus \{j\}} U_i(x_h^{\prime},y_i^{\prime})+ U_j(x_j^{\prime},y_j^{\prime}) \geq \sum_{i\in N} U_i(x_h^{fb},y_i^{fb}),
\end{eqnarray}
where $x_h^{\prime}$ and $y_i^{\prime}$ are the consumption of public and private good, respectively, in the star once $j$ is removed, while $x_j^{\prime}$ and $y_j^{\prime}$ are $j$'s public and private good consumption in isolation, respectively. While the maximization problem is written analogously, in the constraint $x_j^{\prime}$ is added on the LHS, and $(n-2)k$ is subtracted on its RHS. Player $j$'s marginal utilities with respect to the private and public good both equal $\lambda$. Suppose not, then giving more (less) income to $j$ increases social welfare since her marginal utility is larger (smaller) than $\lambda$. If the linking cost is higher than any player's valuation of the public good, this procedure leads to an empty network. This concludes the proof of Proposition \ref{efficiency}. \hfill {$\blacksquare$}

\bigskip

\noindent \textbf{Proof of Corollary \ref{eff_homo}.} Efficient consumption solves \ref{maxeffcons}.
When players have identical preferences, for all $i\in \mathcal{P}$ this yields:
\begin{eqnarray*}
\sum_{i \in g^{star}} \frac{\partial U (x_h^{fb},y_i^{fb})}{\partial x}= \frac{\partial U (x_h^{fb},y_i^{fb})}{\partial y}.
\end{eqnarray*}
Since the LHS is identical for all players, their consumption of the public good is identical. Then, as $U$ is strictly quasi-concave, the planner allocates the same budget for private consumption to all players. This concludes the proof of Corollary \ref{eff_homo}.\hfill {$\blacksquare$}

\bigskip

\noindent \textbf{Proof of Proposition \ref{improving_welfare}.} 
Given $(x^\ast,y^\ast,g^\ast)$, we apply a procedure that yields an equilibrium with a star network in which welfare is higher than in the initial equilibrium. Let player $h$ be such that $\max_{i \in \mathcal{C}(g^\ast)} x_i^\ast \equiv x_h^\ast.$ For any $j\in \mathcal{C}(g^\ast)$ and $j \neq h$, transfer $x^\ast_j$ from $j$ to $h$. For $h$, $\overline{x}^\ast_h=\gamma_h(w_h-(|\mathcal{C}(g^\ast)|-1) k+ \overline{x}^\ast_{-h})$. 
Hence, $h$'s net social income and her demand for the public good are unchanged. Since $\gamma_h'>0$, $h$'s provision increases to $x^\prime_h=\overline{x}^\ast_h$, i.e., $h$ increases her provision precisely by the amount received from the other members of the core.\\
All other players in the core receive an identical amount of public good as before, and any periphery player the same only if she was linked to all core players, and otherwise strictly more. Hence, by Lemma \ref{ind_provision}, each of them still produces less than $k$, and remains in the periphery.\\
Sever any $j$'s link to $i\in\mathcal{C}(\overline{g}^\ast) \setminus \{h\}$, and transfer to player $h$ the amount saved in linking costs $K = k\sum_{i\in N} \sum_{j \neq h} g^\ast_{ij}$. Hence, $w_h^\prime = w_h+\sum_{i \in \mathcal{C}(g^\ast), i\neq h} x^\ast_i + K$, where $w_h$ is $h$'s initial income in the equilibrium $(x^\ast,y^\ast,g^\ast)$.

\noindent Denote the new income distribution $w^{\prime}$ as follows: $w_i^{\prime}=w_i+t_i$ where $t_h=\sum_{i \in N \setminus \{h\}} x^\ast_i + K$, $t_i= -x_i^\ast- k \sum_{j \neq h} g^\ast_{ij}$ for all $j\in\mathcal{C}(g^\ast)$ and $i\neq h$ and $t_p=-k \sum_{p \neq h} g^\ast_{pi}$ for all $p\in\mathcal{P}(g^\ast)$. Hence, $w^{\prime}$ yields a new equilibrium $(x^{\prime},y^{\prime},g^{star})$ where $g^{star}$ is a star network with $h$ as a hub and:

\noindent \textit{(i)} $x_h^{\prime}=\gamma_h(w_h^\prime)\geq \overline{x}^\ast_h$, as $\overline{w}_h^{\prime}\geq \overline{w}_h^{\ast}$ (with $>$ if $g^{\ast}\neq g^{star}$);

\noindent \textit{(ii)} $x_i^{\prime}=0$ and $\overline{x}_i^{\prime} = x_h^\prime \geq \overline{x}_i^\ast$ (with $>$ if $g^{\ast}\neq g^{star}$) for all $i\in \mathcal{C}(g^{\ast})$ and $i\neq h$;

\noindent \textit{(iii)} $x_j^\prime \leq x_j^\ast$ and $\overline{x}_j^{\prime} \geq \overline{x}_j^\ast$ (with strict inequalities if $g^{\ast}\neq g^{star}$) for all $j \in \mathcal{P} (\overline{g}^\ast)$.

\noindent Hence, both welfare and public good consumption are higher in $(x^\prime, y^\prime, g^{star})$ than in $(x^\ast,y^\ast,g^\ast)$. Additionally, welfare in $(x^\prime, y^\prime, g^{star})$ is higher than in any equilibrium with a different network. Suppose not; then, take any other nested split graph $g^{\prime\prime}$ with $h\in\mathcal{C}(g^{\prime\prime})$. Then, we can apply the procedure just described to convert $g^{\prime\prime}$ into a star with $h$ as the hub and associated with a higher welfare, a contradiction.

\noindent This concludes the proof of Proposition \ref{improving_welfare}.\hfill $\blacksquare$

\bigskip

\noindent \textbf{Proof of Proposition \ref{2ndBest}.} By Proposition \ref{improving_welfare}, the social planner wants to achieve a star network. 
Given the total income available $W = \sum_{i \in N} w_i$, denote by $g^{h}$ the star with player $h$ as hub, $\omega^h_i \in [0,1]$ the share of total wealth $W$ assigned to each $i\in N$ such that $\sum_{i\in N}\omega^h_i=1$, by $\mathcal{I}(g^h)$ the set of isolated players in $g^h$, and by $(x^{h},y^{h})$ players' associated consumption choices. 
The social planner solves:
	\begin{eqnarray}
	&& \max_{\omega^h \in [0,1]^{N}, \text{ }h\in N} \sum_{i \in N} U_i (x^h_i, y^h_i)
	\label{second_best}
	\\
	\text{s.t.} && x^h_h =\gamma_h(\omega^h_h W), \label{constraint1a} \\
	&& x^h_h \geq k, \label{constraint1b} \\ 
	&& y^h_h= \omega^h_h W- \gamma_h(\omega^h_h W), 
	\label{constraint1c} \\
	&& x^h_j = \max\{\gamma_j(\omega^h_j W-k+x^h_h)-x^h_h,0\} \text{ for all } j\in\mathcal{P}(g^h),
	\label{constraint2a}\\
	&& x^h_j \leq k \text{ for all } j \in\mathcal{P}(g^h), \label{constraint2b} \\
	&& y_j^h =\omega^h_j W -k-x^h_j \text{ for all } j \in\mathcal{P}(g^h), \label{constraint2c} \\
	&& U_i (x^I_i, y^I_i)\geq U_i (x+x^h_h, \omega^h_i W -k-x), x=\max\{\gamma_i(\omega^h_i W-k+x^h_h)-x^h_h,0\} \notag \\
	&& \text{ for all } i\in \mathcal{I}(g^h). \label{constraint3}
\end{eqnarray}
Abstracting for a moment from the inequality constraints and assuming that the changes in $\omega_i^h$ do not turn $i$ into an active player if inactive (or vice versa), the FOC with respect to $\omega_i^h$ for $i\in \mathcal{P}(g^h)$ yields
\begin{eqnarray}\label{foc_sb}
	&&\frac{\partial U_h}{\partial x}\gamma_h^{\prime}+\frac{\partial U_h}{\partial y}\frac{1-\gamma_h^{\prime}}{p_h}+ \notag \\ &+& 
	\gamma^{\prime}_h
	\left[\sum_{j \in \mathcal{P}(g^h) \setminus \{i\} | x_j=0 } \frac{\partial U_j}{\partial x} +\sum_{j \in \mathcal{P}(g^h) \setminus \{i\} | x_j>0
	} \left( \frac{\partial U_j}{\partial x} \gamma_j^{\prime} + \frac{\partial U_j}{\partial y}\frac{1-\gamma^{\prime}_j}{p_j}\right) \right]= \notag \\=
	&I_{x_i=0}& \left(-\frac{\partial U_i}{\partial x} \gamma_h^{\prime} +\frac{\partial U_i}{\partial y}\frac{1}{p_i} \right)+
	I_{x_i>0} (1-\gamma^{\prime}_h) \left( \frac{\partial U_i}{\partial x} \gamma_i^{\prime} 
	+ \frac{\partial U_i}{\partial y}\frac{(1-\gamma_i^{\prime})}{p_i} \right),
\end{eqnarray}
where $I_{x_i>0}$ ($I_{x_i=0}$) in an indicator function equal to $1$ if $x_i>0$ ($x_i=0$), and zero otherwise, while we write $\gamma_i$ instead of $\gamma_i(\tilde{w}_i)$ for all $i\in N$. Condition \eqref{foc_sb} shows the different effects of distributing resources from a player $i$ to the hub $h$. On the one hand, the LHS describes the increase in consumption of the hub, as well as the increase in public good consumption of periphery players different from $i$, whose public good consumption increases, while potentially decreasing their provision, hereby also increasing their private consumption. On the other hand, the RHS shows the effect on player $i$. If $i$ is inactive, the public good $i$ accesses from $h$ decreases, and the income is used for the private good. If $i$ is active, $i$'s contribution to the public good increases, implying a lower increase in private good consumption. The planner chooses the allocation that balances these trade-offs. Let us denote by $\tilde{w}$ the wealth distribution induced by the weights derived using these FOCs.

\noindent Consider now $g^h$ were player $h$ is the player with most links in $(x^{\ast},y^{\ast},g^{\ast})$. A solution to \ref{second_best} fixing $g^h$ exists if $\tilde{w}$ is such that \eqref{constraint1c}, \eqref{constraint2c} and \eqref{constraint3} are satisfied. In other words, $\tilde{w}$ must be such that the hub of the star produces more than $k$, players in the periphery less than $k$ and isolated players do not want to join the star. Note that \eqref{constraint1c} and \eqref{constraint3} are satisfied by construction: if \eqref{constraint1c} was not satisfied, then the planner would implement the empty network, while if a player values the public good enough, she is part of the star. Condition \eqref{constraint2c} is satisfied for sure if $\tilde{w}_h\geq w_h+\sum_{i\in\mathcal{C}(g^\ast)} x^\ast_i$, as all periphery players produce less than $k$ of public good by construction. Otherwise, using the same arguments as in the proof of Proposition \ref{existence}, there exists a $\Gamma$ such that, for all $i\neq h$, if $\gamma^{\prime}_i<\Gamma$, \eqref{constraint2c} holds. Then, \eqref{second_best} admits at least a solution.

\noindent Finally, the planner derives the weights solving \ref{second_best} for each player $i\in N$ as a hub, and compares welfare between these star networks. Since at least one solution exists (for $g^h$), and the set of potential hubs is finite, a solution exists. This concludes the proof 
of Proposition \ref{2ndBest}. \hfill {$\blacksquare $}

\bigskip

\noindent \textbf{Proof of Corollary \ref{linear}.} Denote by $\beta \in (0,1)$ the share of total wealth assigned to the hub, by $p$ any periphery player in $\mathcal{P}(\overline{g}^\prime)$ and by $(x^{sb},y^{sb},g^{star})$ the second-best allocation. Given $g^\prime$, the social planner chooses $\beta^{sb}$ to maximize welfare:
	\begin{eqnarray}
	\max_{\beta \in (0,1)} && h(\alpha) \beta W +f(\alpha)+ I_{x_p=0}(n- 1)U(x_h, \frac{(1-\beta)W}{n-1}-k) \notag \\ &&+ I_{x_p>0} [h(\alpha) ((1-\beta) W-k(n- 1))+(n- 1) f(\alpha)]
	\label{secondbestbeta}\\
	\text{s.t. } && x_h=\alpha \beta W\geq k \label{c1}\\
	&& \frac{(1-\beta) W}{n-1}\geq k \label{c2} \\
	&& \alpha \frac{(1-\beta) W}{n-1}-k-(1-\alpha) \beta W \leq k \label{c3}
	\end{eqnarray}
where $h(\alpha) \beta W +f(\alpha)$ is the associated indirect utility linear in the hub's income $\beta W$; $h$ and $f$ depend on the slope of the Engel curve $\alpha$ (and the price of the public good, but this is the same for all active players, so we suppress this argument). \eqref{c1} ensures the hub produces more than the linking cost, \eqref{c2} that periphery players have enough resources to link to the hub and \eqref{c3} that they do not produce more than the linking cost. As the objective function is continuous in $\beta$ and we are maximizing over a compact set, existence of a solution follows. We now turn to characterizing the solution.

\noindent If $y^{sb}_i=0$ for all $i\in N$, $\alpha=1$ and $U_2(\cdot,\cdot)=0$. Then, \eqref{secondbestbeta} simplifies to
\[
\max_{\beta \in (0,1)} h(\alpha) \beta W + f(\alpha)+ (n- 1)U( \beta W-k+\frac{(1-\beta)W}{n-1}, 0),
\]
leading to the FOC $h(\alpha) +(n-2) U_1(\beta W-k+(1-\beta)W/(n-1), 0 ) = 0$. Since $U_1(\cdot,\cdot)$ is positive, if $n>2$, $\beta^{sb}=1$.

\noindent Suppose $x_i^{sb},y_i^{sb}>0$ for all $i\in N$. Then, we would get $\beta^{sb}$ such that $\alpha \beta^{sb} W=k$ and $\alpha [(1-\beta^{sb})W/(n-1)-k]<k$. So, this is a solution if $\alpha<1/(n-1)$ and $W<k(\alpha(n-1)+n)/\alpha$.

\noindent Finally, if $y_i^{sb}>0$ for all $i\in N$ and $x_i^{sb}>0$ only for $h$, we obtain
\[
h(\alpha) + (n- 1) \alpha U_1(\alpha \beta W, \frac{(1-\beta)W}{n-1}-k)=U_2(\alpha \beta W, \frac{(1-\beta)W}{n-1}-k).
\]
Hence, an increase in $\beta$ increases the hub's public and private good consumption and the public good consumption of the periphery players, while it decreases their private good consumption. The social planner chooses $\beta^{sb}$ to balance these two effects. This concludes the proof of Corollary \ref{linear}. \hfill $\blacksquare$

\bigskip

\noindent \textbf{Proof of Proposition \ref{personalized prices}.} Without loss of generality, assume that $p_i=1$ for all $i\in N$. Suppose first that all players belong to a star. In the efficient solution, the social planner's budget constraint $\sum_{i \in N} w_i = x_h + \sum_{i \in N} y_i + (n-1)k$ holds. Choose the personalized price $p_x$ for the hub, without loss of generality, player $h$, such that $\left(\partial U_h(x_h, y_h) \slash \partial x \right)/\left(\partial U_h(x_h, y_h) \slash \partial y \right) = p_x$, as $\gamma_h^\prime \in (0,1)$. 
\noindent Players $i \neq h$, link to $h$ and provide no public good since, by Proposition \ref{efficiency}, $\partial U_i(x_h, y_i) \slash \partial x <\partial U_i(x_h, y_i) \slash \partial y_i$.
The social planner charges taxes $\tau,$ such that $\tau_i > 0$ for all $i \neq h,$ to reduce player $h$'s price of the public good by $(1-p_x) x_h.$ Given $w,$ choose $\tau_h$ such that $(w_h - \tau_h) - y_h^{fb} - x_h^{fb} p_x = 0,$ and for $i \neq h,$ choose $\tau_i$ such that $(w_i - \tau_i) - y_i^{fb} - k = 0.$ This implements the efficient provision of Proposition \ref{efficiency}. To see that the solution is budget balanced, sum the budget constraints of all players $\sum_{i \in N} (w_i - \tau_i) = x_h^{fb} p_x + \sum_{i \in N} y_i^{fb} + (n-1) k,$ 
and substitute for $\sum_{i \in N} w_i$ from the social planner's budget constraint in the efficient solution. This yields $\sum_{i \in N} \tau_i = (1-p_x)x_h^{fb}$ as required.

\noindent If some players are isolated in the efficient solution of Proposition \ref{efficiency}, the provision of the star's hub is obtained as above, and for any isolated player $j$, $\tau_j$ is chosen such that $(w_j - \tau_j) - x_j^{\prime} - y_j^{\prime} = 0,$ where $x_j^{\prime}$ and $y_j^{\prime}$ are obtained from \eqref{obf2}. This concludes the proof of Proposition \ref{personalized prices}. \hfill {$\blacksquare$}

\bibliographystyle{elsarticle-harv}
\bibliography{library-EER}

\begin{thebibliography}{26}
\expandafter\ifx\csname natexlab\endcsname\relax\def\natexlab#1{#1}\fi
\providecommand{\url}[1]{\texttt{#1}}
\providecommand{\href}[2]{#2}
\providecommand{\path}[1]{#1}
\providecommand{\DOIprefix}{doi:}
\providecommand{\ArXivprefix}{arXiv:}
\providecommand{\URLprefix}{URL: }
\providecommand{\Pubmedprefix}{pmid:}
\providecommand{\doi}[1]{\href{http://dx.doi.org/#1}{\path{#1}}}
\providecommand{\Pubmed}[1]{\href{pmid:#1}{\path{#1}}}
\providecommand{\bibinfo}[2]{#2}
\ifx\xfnm\relax \def\xfnm[#1]{\unskip,\space#1}\fi
\bibitem[{Allouch(2015)}]{nizar}
\bibinfo{author}{Allouch, N.}, \bibinfo{year}{2015}.
\newblock \bibinfo{title}{On the private provision of public goods on
  networks}.
\newblock \bibinfo{journal}{Journal of Economic Theory} \bibinfo{volume}{157},
  \bibinfo{pages}{527--552}.
\bibitem[{Baetz(2015)}]{baetz}
\bibinfo{author}{Baetz, O.}, \bibinfo{year}{2015}.
\newblock \bibinfo{title}{Social activity and network formation}.
\newblock \bibinfo{journal}{Theoretical Economics} \bibinfo{volume}{10},
  \bibinfo{pages}{315--340}.
\bibitem[{Bala and Goyal(2000)}]{balagoyal}
\bibinfo{author}{Bala, V.}, \bibinfo{author}{Goyal, S.}, \bibinfo{year}{2000}.
\newblock \bibinfo{title}{A noncooperative model of network formation}.
\newblock \bibinfo{journal}{Econometrica} \bibinfo{volume}{68},
  \bibinfo{pages}{1181--1229}.
\bibitem[{Ballester et~al.(2006)Ballester, Calv{\'o}-Armengol and
  Zenou}]{keyplayer}
\bibinfo{author}{Ballester, C.}, \bibinfo{author}{Calv{\'o}-Armengol, A.},
  \bibinfo{author}{Zenou, Y.}, \bibinfo{year}{2006}.
\newblock \bibinfo{title}{Who's who in networks. wanted: The key player}.
\newblock \bibinfo{journal}{Econometrica} \bibinfo{volume}{74},
  \bibinfo{pages}{1403--1417}.
\bibitem[{Banks et~al.(1997)Banks, Blundell and Lewbel}]{quadratic}
\bibinfo{author}{Banks, J.}, \bibinfo{author}{Blundell, R.},
  \bibinfo{author}{Lewbel, A.}, \bibinfo{year}{1997}.
\newblock \bibinfo{title}{Quadratic engel curves and consumer demand}.
\newblock \bibinfo{journal}{Review of Economics and Statistics}
  \bibinfo{volume}{79}, \bibinfo{pages}{527--539}.
\bibitem[{Bastos et~al.(2018)Bastos, Piccardi, Levy, McRoberts and
  Lubell}]{bastos}
\bibinfo{author}{Bastos, M.}, \bibinfo{author}{Piccardi, C.},
  \bibinfo{author}{Levy, M.}, \bibinfo{author}{McRoberts, N.},
  \bibinfo{author}{Lubell, M.}, \bibinfo{year}{2018}.
\newblock \bibinfo{title}{Core-periphery or decentralized? topological shifts
  of specialized information on twitter}.
\newblock \bibinfo{journal}{Social Networks} \bibinfo{volume}{52},
  \bibinfo{pages}{282--293}.
\bibitem[{Baumann(2018)}]{baumann2018self}
\bibinfo{author}{Baumann, L.}, \bibinfo{year}{2018}.
\newblock \bibinfo{title}{Self-ratings and peer review}.
\newblock \bibinfo{journal}{Mimeo} .
\bibitem[{Belhaj et~al.(2016)Belhaj, Bervoets and Dero{\"\i}an}]{belhaj2016}
\bibinfo{author}{Belhaj, M.}, \bibinfo{author}{Bervoets, S.},
  \bibinfo{author}{Dero{\"\i}an, F.}, \bibinfo{year}{2016}.
\newblock \bibinfo{title}{Efficient networks in games with local
  complementarities}.
\newblock \bibinfo{journal}{Theoretical Economics} \bibinfo{volume}{11},
  \bibinfo{pages}{357--380}.
\bibitem[{Bergstrom et~al.(1986)Bergstrom, Blume and Varian}]{BBV}
\bibinfo{author}{Bergstrom, T.}, \bibinfo{author}{Blume, L.},
  \bibinfo{author}{Varian, H.}, \bibinfo{year}{1986}.
\newblock \bibinfo{title}{On the private provision of public goods}.
\newblock \bibinfo{journal}{Journal of Public Economics} \bibinfo{volume}{29},
  \bibinfo{pages}{25--49}.
\bibitem[{Billand et~al.(2008)Billand, Bravard and
  Sarangi}]{billand2008existence}
\bibinfo{author}{Billand, P.}, \bibinfo{author}{Bravard, C.},
  \bibinfo{author}{Sarangi, S.}, \bibinfo{year}{2008}.
\newblock \bibinfo{title}{Existence of nash networks in one-way flow models}.
\newblock \bibinfo{journal}{Economic Theory} \bibinfo{volume}{37},
  \bibinfo{pages}{491--507}.
\bibitem[{Bloch and Olckers(forthcoming)}]{bloch2021friend}
\bibinfo{author}{Bloch, F.}, \bibinfo{author}{Olckers, M.},
  \bibinfo{year}{forthcoming}.
\newblock \bibinfo{title}{Friend-based ranking}.
\newblock \bibinfo{journal}{American Economic Journal: Microeconomics} .
\bibitem[{Bramoull{\'e} and Kranton(2007)}]{brkran}
\bibinfo{author}{Bramoull{\'e}, Y.}, \bibinfo{author}{Kranton, R.},
  \bibinfo{year}{2007}.
\newblock \bibinfo{title}{Public goods in networks}.
\newblock \bibinfo{journal}{Journal of Economic Theory} \bibinfo{volume}{135},
  \bibinfo{pages}{478--494}.
\bibitem[{Bramoull{\'e} et~al.(2014)Bramoull{\'e}, Kranton and
  D'Amours}]{brkrandam}
\bibinfo{author}{Bramoull{\'e}, Y.}, \bibinfo{author}{Kranton, R.},
  \bibinfo{author}{D'Amours, M.}, \bibinfo{year}{2014}.
\newblock \bibinfo{title}{Strategic interaction and networks}.
\newblock \bibinfo{journal}{The American Economic Review}
  \bibinfo{volume}{104}, \bibinfo{pages}{898--930}.
\bibitem[{Conley and Udry(2010)}]{udry}
\bibinfo{author}{Conley, T.}, \bibinfo{author}{Udry, C.}, \bibinfo{year}{2010}.
\newblock \bibinfo{title}{Learning about a new technology: Pineapple in ghana}.
\newblock \bibinfo{journal}{The American Economic Review}
  \bibinfo{volume}{100}, \bibinfo{pages}{35--69}.
\bibitem[{Elliott and Golub(2019)}]{golub}
\bibinfo{author}{Elliott, M.}, \bibinfo{author}{Golub, B.},
  \bibinfo{year}{2019}.
\newblock \bibinfo{title}{A network approach to public goods}.
\newblock \bibinfo{journal}{Journal of Political Economy}
  \bibinfo{volume}{127}, \bibinfo{pages}{730--776}.
\bibitem[{Feick and Price(1987)}]{feick}
\bibinfo{author}{Feick, L.}, \bibinfo{author}{Price, L.}, \bibinfo{year}{1987}.
\newblock \bibinfo{title}{The market maven: A diffuser of marketplace
  information}.
\newblock \bibinfo{journal}{The Journal of Marketing} \bibinfo{volume}{51},
  \bibinfo{pages}{83--97}.
\bibitem[{Fershtman and Persitz(2021)}]{dotan}
\bibinfo{author}{Fershtman, C.}, \bibinfo{author}{Persitz, D.},
  \bibinfo{year}{2021}.
\newblock \bibinfo{title}{Social clubs and social networks}.
\newblock \bibinfo{journal}{American Economic Journal: Microeconomics}
  \bibinfo{volume}{13}, \bibinfo{pages}{224--251}.
\bibitem[{Galeotti(2006)}]{galeotti2006one}
\bibinfo{author}{Galeotti, A.}, \bibinfo{year}{2006}.
\newblock \bibinfo{title}{One-way flow networks: The role of heterogeneity}.
\newblock \bibinfo{journal}{Economic Theory} \bibinfo{volume}{29},
  \bibinfo{pages}{163--179}.
\bibitem[{Galeotti and Goyal(2010)}]{gg}
\bibinfo{author}{Galeotti, A.}, \bibinfo{author}{Goyal, S.},
  \bibinfo{year}{2010}.
\newblock \bibinfo{title}{The law of the few}.
\newblock \bibinfo{journal}{The American Economic Review}
  \bibinfo{volume}{100}, \bibinfo{pages}{1468--1492}.
\bibitem[{Herskovic and Ramos(2020)}]{ramos}
\bibinfo{author}{Herskovic, B.}, \bibinfo{author}{Ramos, J.},
  \bibinfo{year}{2020}.
\newblock \bibinfo{title}{Acquiring information through peers}.
\newblock \bibinfo{journal}{The American Economic Review}
  \bibinfo{volume}{110}, \bibinfo{pages}{2128--52}.
\bibitem[{Hiller(2017)}]{hiller}
\bibinfo{author}{Hiller, T.}, \bibinfo{year}{2017}.
\newblock \bibinfo{title}{Peer effects in endogenous networks}.
\newblock \bibinfo{journal}{Games and Economic Behavior} \bibinfo{volume}{105},
  \bibinfo{pages}{349--367}.
\bibitem[{Kinateder and Merlino(2017)}]{KM}
\bibinfo{author}{Kinateder, M.}, \bibinfo{author}{Merlino, L.P.},
  \bibinfo{year}{2017}.
\newblock \bibinfo{title}{Public goods in endogenous networks}.
\newblock \bibinfo{journal}{American Economic Journal: Microeconomics}
  \bibinfo{volume}{9}, \bibinfo{pages}{187--212}.
\bibitem[{K{\"o}nig et~al.(2014)K{\"o}nig, Tessone and Zenou}]{ktz}
\bibinfo{author}{K{\"o}nig, M.D.}, \bibinfo{author}{Tessone, C.J.},
  \bibinfo{author}{Zenou, Y.}, \bibinfo{year}{2014}.
\newblock \bibinfo{title}{Nestedness in networks: A theoretical model and some
  applications}.
\newblock \bibinfo{journal}{Theoretical Economics} \bibinfo{volume}{9},
  \bibinfo{pages}{695--752}.
\bibitem[{Van~Leeuwen et~al.(2019)Van~Leeuwen, Offerman and
  Schram}]{leeuwenJEEA}
\bibinfo{author}{Van~Leeuwen, B.}, \bibinfo{author}{Offerman, T.},
  \bibinfo{author}{Schram, A.}, \bibinfo{year}{2019}.
\newblock \bibinfo{title}{{Competition for Status Creates Superstars: An
  Experiment on Public Good Provision and Network Formation}}.
\newblock \bibinfo{journal}{Journal of the European Economic Association}
  \bibinfo{note}{Jvz001}.
\bibitem[{Warr(1983)}]{warr}
\bibinfo{author}{Warr, P.G.}, \bibinfo{year}{1983}.
\newblock \bibinfo{title}{The private provision of a public good is independent
  of the distribution of income}.
\newblock \bibinfo{journal}{Economics Letters} \bibinfo{volume}{13},
  \bibinfo{pages}{207--211}.
\bibitem[{Zhang et~al.(2007)Zhang, Ackerman and Adamic}]{adamic}
\bibinfo{author}{Zhang, J.}, \bibinfo{author}{Ackerman, M.S.},
  \bibinfo{author}{Adamic, L.}, \bibinfo{year}{2007}.
\newblock \bibinfo{title}{Expertise networks in online communities: Structure
  and algorithms}.
\newblock \bibinfo{journal}{Proceedings of the 16th International Conference on
  World Wide Web} , \bibinfo{pages}{221--230}.

\end{thebibliography}

\end{document}